\documentclass{emulateapj}

\usepackage{multirow}
\usepackage{amsmath}
\usepackage{longtable}
\usepackage{soul}

\newcommand{\twco} {^{12}\mbox{CO}}
\newcommand{\thco} {^{13}\mbox{CO}}
\newcommand{\ceio} {\mbox{C}^{18}\mbox{O}}
\newcommand{\cseo} {\mbox{C}^{17}\mbox{O}}
\newcommand{\cths} {\mbox{C}^{34}\mbox{S}}
\newcommand{\Htwo} {\mbox{H}_2}
\newcommand{\h} {^{\mathrm{h}}}
\newcommand{\m} {^{\mathrm{m}}}
\newcommand{\s} {^{\mathrm{s}}}

\newcommand{\Tex}       {T_{\mathrm{ex}}}
\newcommand{\Tbg}       {T_{\mathrm{bg}}}
\newcommand{\Qrot}      {Q_{\mathrm{rot}}}
\newcommand{\vout}     {v_\mathrm{out}}
\newcommand{\Mout}     {M_\mathrm{out}}
\newcommand{\Pout}     {P_\mathrm{out}}
\newcommand{\Eout}     {E_\mathrm{out}}

\newcommand{\Jybeam}  {\mbox{Jy}~\mbox{beam}^{-1}}
\newcommand{\mJybeam}  {\mbox{mJy}~\mbox{beam}^{-1}}
\newcommand{\Jybeamkms}  {\mbox{Jy}~\mbox{beam}^{-1}~\mbox{km s}^{-1}}

\newcommand{\cm}	{\mbox{cm}}

\newcommand{\kms}	{\mbox{km s}^{-1}}

\newcommand{\yr}	{{\rm yr}}

\newcommand{\aukms} {\mbox{AU km s}^{-1}}

\shorttitle{ALMA Cycle 1 Observations of the HH46/47 Molecular Outflow}
\shortauthors{Zhang et al.}

\begin{document}

\title{ALMA Cycle 1 Observations of the HH46/47 Molecular Outflow: Structure, Entrainment and Core Impact}

\author{Yichen Zhang$^{1,2}$, H\'ector G. Arce$^2$, Diego Mardones$^1$, Sylvie Cabrit$^{3,4}$,
Michael M. Dunham$^5$, Guido Garay$^1$, Alberto Noriega-Crespo$^6$, Stella S. R. Offner$^7$,
Alejandro C. Raga$^8$, Stuartt A. Corder$^9$}

\affil{$^1$Departamento de Astronom\'ia, Universidad de Chile, Casilla 36-D, Santiago, Chile\\
$^2$Astronomy Department, Yale University, P.O. Box 208101, New Haven, CT 06520, USA\\
$^3$LERMA, Observatoire de Paris, UMR 8112 du CNRS, ENS, UPMC, UCP, 61 Av. de l'Observatoire, F-75014 Paris, France\\
$^4$Institut de Plan\'etologie et d'Astrophysique de Grenoble (IPAG) UMR 5274, Grenoble, 38041, France\\
$^5$Harvard-Smithsonian Center for Astrophysics, 60 Garden Street, Cambridge, MA 02138, USA\\
$^6$Space Telescope Science Institute, 3700 San Martin Dr., Baltimore, MD 21218, USA\\
$^7$Department of Astronomy, University of Massachusetts, Amherst, MA 01002, USA\\
$^8$Instituto de Ciencias Nucleares, UNAM, Ap. 70-543, 04510 D.F., Mexico\\
$^9$Joint ALMA Observatory, Av. Alonso de C\'ordova 3107, Vitacura, Santiago, Chile\\
yczhang.astro@gmail.com}

\begin{abstract}
We present Atacama Large Millimeter/sub-millimeter Array 
Cycle 1 observations of the HH 46/47 molecular outflow 
using combined 12m array and Atacama Compact Array observations. 
The improved angular resolution and sensitivity of our multi-line maps 
reveal structures that help us study the entrainment process in much more 
detail and allow us to obtain more precise estimates of outflow properties 
than previous observations.
We use $\thco$ (1-0) and $\ceio$ (1-0) emission to correct for the $\twco$ (1-0) optical depth 
to accurately estimate the outflow mass, momentum and kinetic energy. 
This correction increases the estimates of the mass, momentum and kinetic energy by 
factors of about 9, 5 and 2, respectively, with respect to estimates assuming optically thin emission. 
The new $\thco$ and $\ceio$ data also allow us to trace denser and slower outflow material than that 
traced by the $\twco$ maps, and they reveal an outflow cavity wall at very low velocities 
(as low as $0.2~\kms$ with respect to the coreÕs central velocity). 
Adding with the slower material traced only by $\thco$ and $\ceio$, 
there is another factor of 3 increase in the mass estimate and 50\% increase in the momentum estimate. 
The estimated outflow properties indicate that the outflow is capable of dispersing the parent core 
within the typical lifetime of the embedded phase of a low-mass protostar, 
and that it is responsible for a core-to-star efficiency of 1/4 to 1/3.  
We find that the outflow cavity wall is composed of multiple shells associated with a series of jet bow-shock events.
Within about 3000 AU of the protostar the $\thco$ and $\ceio$ emission trace a circumstellar envelope 
with both rotation and infall motions, which we compare with a simple analytic model.  
The CS (2-1) emission reveals tentative evidence of a slowly-moving rotating outflow, 
which we suggest is entrained not only poloidally but also toroidally by a disk wind that 
is launched from relatively large radii from the source.
\end{abstract}

\keywords{ISM: clouds, Herbig-Haro objects, individual objects (HH 46, HH 47), jets and outflows 
--- stars: formation}

\section{Introduction}
\label{sec:intro}

Outflows play an important role in star formation and the evolution of molecular clouds and cores.
They carve out cavities in their parent cores and
inject energy and momentum into the star-forming environment. 
They may be responsible for dispersing the core (\citealt[]{Arce06}), 
terminating the infall phase (e.g., \citealt[]{Velusamy98}), and thereby 
determining the final stellar mass and the core-to-star efficiency (e.g., \citealt[]{Matzner00};
\citealt[]{Myers08}; \citealt[]{Offner14a}). In particular, a
nearly constant 30\% efficiency due to outflows would be one explanation for the similar shape of
the core mass function (CMF) and the initial mass function (IMF) (e.g., \citealt[]{Alves07}; \citealt[]{Offner14}).
However, it is still unclear whether the outflow is powerful enough to disperse 70\% of the
gas in the surrounding core. While some studies have shown that outflows have a profound
effect on the environment surrounding the protostar and are able to disperse the parent core
on timescales less than 1 Myr (e.g., \citealt[]{Tafalla97}; \citealt[]{Fuller02}; \citealt[]{Arce06}),
other studies have claimed that the mass-loss rate from the outflows is too low and outflows cannot be the
sole agent responsible for core dispersal (e.g., \citealt[]{Hatchell07}; \citealt[]{Curtis10}).
More studies with reliable estimates of the outflow mass, momentum and energy are needed to solve this discrepancy.
Considering that the denser material at low velocities, which is untraceable by optically thick $\twco$
emission, may contribute a large fraction of the outflow mass, optically thinner tracers like $\thco$
and $\ceio$ are needed to understand the impact of the outflow on the denser material (\citealt[]{Arce06}). 

The accretion of material from the circumstellar disk onto the protostar drives bipolar magneto-centrifugal winds.
A collimated wind or the collimated portion of a wind, which is typically observed in atomic lines,
is usually referred to as a jet. 
The molecular outflow is believed to be the ambient gas entrained by such bipolar winds.
The entrainment process is not yet clearly understood.
Models include entrainment through wide-angle winds (e.g. \citealt[]{Li96}) and jet bow-shocks (internal
and/or leading) (e.g. \citealt[]{Raga93}).
In the wide-angle wind entrainment model, a radial wind blows into the ambient material, forming a thin
outflowing shell. In the jet bow-shock entrainment model, a jet propagates into the ambient material and
forms bow-shocks which accelerate the ambient gas producing outflow shells surrounding the jet.
These two mechanisms may co-exist but one may play a more
important role than the other depending on the distribution of the ambient material.

This paper is a follow-up study of the HH 46/47 molecular outflow (\citealt[]{Arce13}, Paper I hereafter), 
using Atacama Large Millimeter/sub-millimeter Array (ALMA) Cycle 1 observations.
The HH 46/47 outflow is driven by a low-mass early Class I protostar (HH 47 IRS, HH 46 IRS 1, IRAS 08242-5050, 
$12~L_\odot$) which resides in the Bok globule ESO 216-6A, located on the outskirts of the 
Gum Nebula at a distance of 450 pc (\citealt[]{Schwartz77}; \citealt[]{Reipurth00}; \citealt[]{Noriega04}).
{\it Hubble Space Telescope} ({\it HST}) observations indicate that HH 47 IRS is 
a binary system with an observed separation between the two components of the system of
 $0.\arcsec26$ or about 120 AU (\citealt[]{Reipurth00}).
As the driving source lies very close to the edge of the globule, the blue-shifted outflow
can be seen at optical wavelengths extending outside of the globule, while most of 
the red-shifted outflow lies inside of the globule and therefore is best seen in infrared.
Wide-field, narrowband H$\alpha$ and [S II] optical images by \citet[]{Stanke99} revealed
that the jet may extend further away from the globule, extending 2.6 pc on the plane of the sky.
The HH 46/47 outflow has been extensively studied in optical and infrared.
Studies combining the optical spectral data and proper motion observations estimated 
the average jet velocity to be 300 $\kms$ and the inclination between the jet and the plane of the sky
to be about $30^\circ$ to $40^\circ$ (\citealt[]{Reipurth89}; \citealt[]{Reipurth91}; \citealt[]{Eisloffel94};
\citealt[]{Micono98}; \citealt[]{Hartigan05}). In addition, the infrared shocked $\Htwo$ emission
also shows an outflow cavity structure with a width of 36$\arcsec$ and a length of about $2\arcmin$ 
(\citealt[]{Eisloffel94a}; \citealt[]{Noriega04}).

The HH 46/47 molecular outflow was recently observed in $\twco$ (1-0) by ALMA in Cycle 0
(Paper I). The molecular outflow appears to be highly asymmetric: the blue-shifted lobe
extends no more than 30$\arcsec$, while the red-shifted lobe extends about $2\arcmin$. 
Detailed analysis of the morphology and kinematics of the molecular outflow showed evidence
of wide-angle wind entrainment for the blue-shifted outflow and jet bow-shock entrainment for the 
red-shifted outflow.
These asymmetries are due to the fact that the blue-shifted jet is mostly outside of the globule where there is
little or no molecular gas for it to entrain, while the red-shifted jet plunges into the parent cloud.  
APEX and Herschel observations revealed the existence of warm CO, H$_2$O, OH, and [O I] emission
in this outflow, produced by shocks where the protostellar jet/wind interacts with the parent core 
(\citealt[]{vanKempen09}; \citealt[]{vanKempen10}; \citealt[]{Wampfler10}).

Here we present analysis of ALMA Cycle 1 observation of $\twco$ (1-0), $\thco$ (1-0), $\ceio$ (1-0), CS (2-1)
and other molecules using the ALMA 12m array and Atacama Compact Array (ACA). 
Molecules such as $\thco$ and $\ceio$ trace higher column density gas than $\twco$, 
allowing us to obtain a more accurate assessment of the outflow's impact on the core.
We also use $\thco$ and $\ceio$ to correct the CO opacity to more accurately determine
the properties of the outflow. 
Compared with previous ALMA observation, the new $\twco$ data also have improved angular resolution 
and sensitivity to large structures, and reveal richer details of the outflow that help constrain the outflow
entrainment mechanism.

\section{Observations}
\label{sec:obser}

\begin{figure*}
\begin{center}
\includegraphics[width=\textwidth]{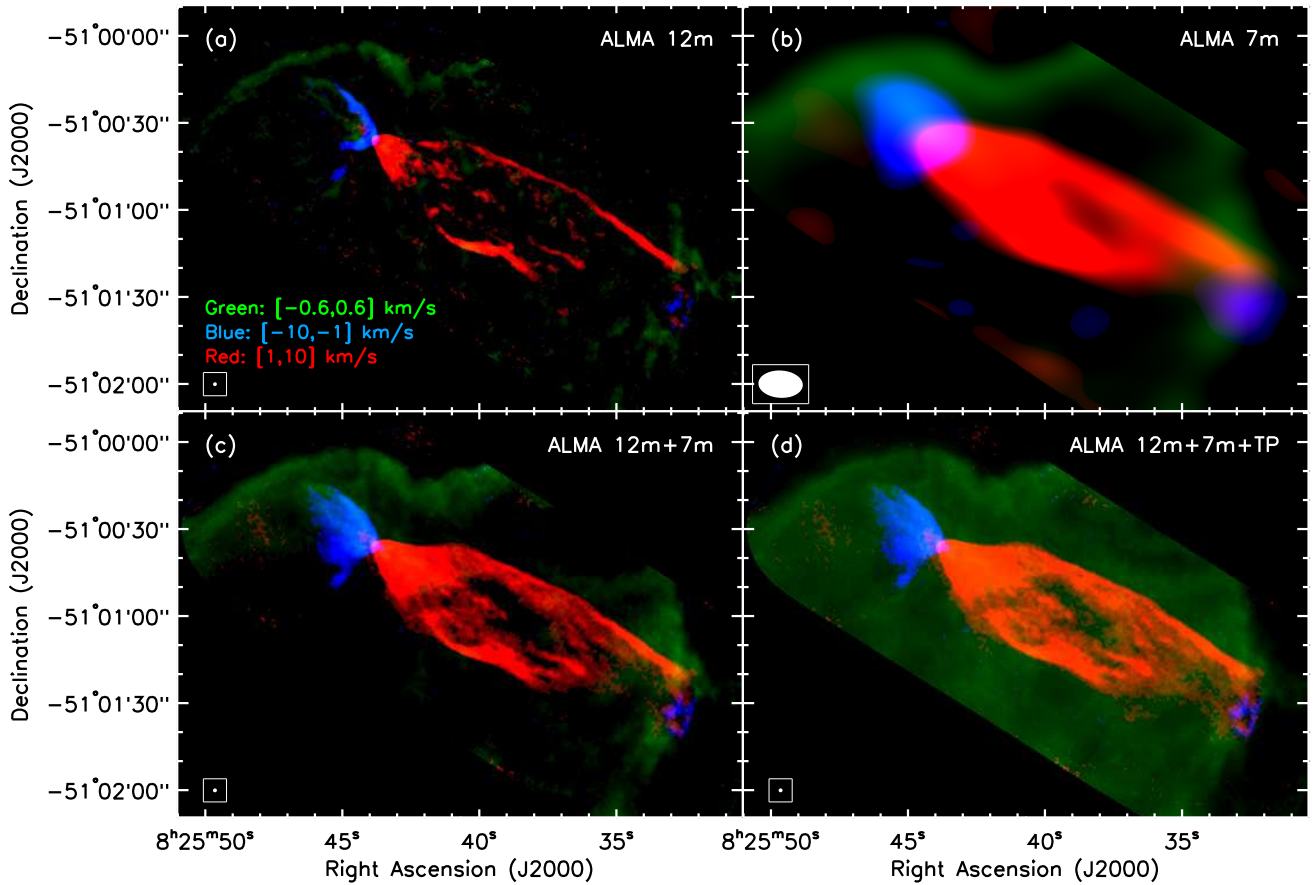}\\
\caption{Three-color images showing $\twco$ (1-0) integrated emission from (a) the 12m array data, 
(b) the 7m array data, (c) the interferometric data combining the 12m array and 7m array, 
and (d) the combined data of the interferometric data and total power data. 
The red, green and blue color scales show emission integrated over the velocity ranges from
1 to $10~\kms$, from $-0.6$ to $0.6~\kms$ and from $-10$ to $-1~\kms$ (relative to the cloud velocity) respectively.
From panel (a) to (d), the synthesized beams are 
$1.33\arcsec \times 1.28\arcsec$ (P.A. = $-59.7^\circ$), $15.3\arcsec \times 9.4\arcsec$ (P.A. = $86.8^\circ$), 
$1.37\arcsec \times 1.31\arcsec$ (P.A. = $-58.2^\circ$), and 
$1.37\arcsec \times 1.31\arcsec$ (P.A. = $-58.2^\circ$) respectively 
(the white ellipse in the lower-left corner of each panel).}
\label{fig:comb}
\end{center}
\end{figure*}

\begin{figure} 
\begin{center}
\includegraphics[width=\columnwidth]{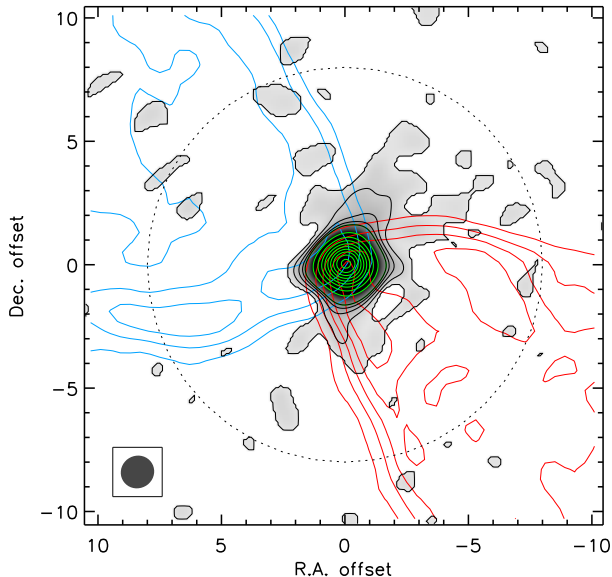}\\
\caption{The 100 GHz continuum map towards the central source.
Only the interferometric data of the two 2-GHz bands are used.
Black contours start at 3$\sigma$ and end at 15$\sigma$ with a step of 3$\sigma$.
Here 1$\sigma=0.041~\mJybeam$ (2.32 mK). 
Green contours start at 30$\sigma$ with a step of 15$\sigma$ to show the high intensity part.
The highest contour level is 150$\sigma$.
The synthesized beam of the continuum map is $1.48\arcsec \times 1.41\arcsec$ with P.A. = -77.3$^\circ$. 
The red and blue contours show red-shifted and blue-shifted $\twco$ outflows for reference.
The dashed circle defines the region we integrate to obtain the total flux.}
\label{fig:continuummap}
\end{center}
\end{figure}

\begin{figure*} 
\begin{center}
\includegraphics[width=\textwidth]{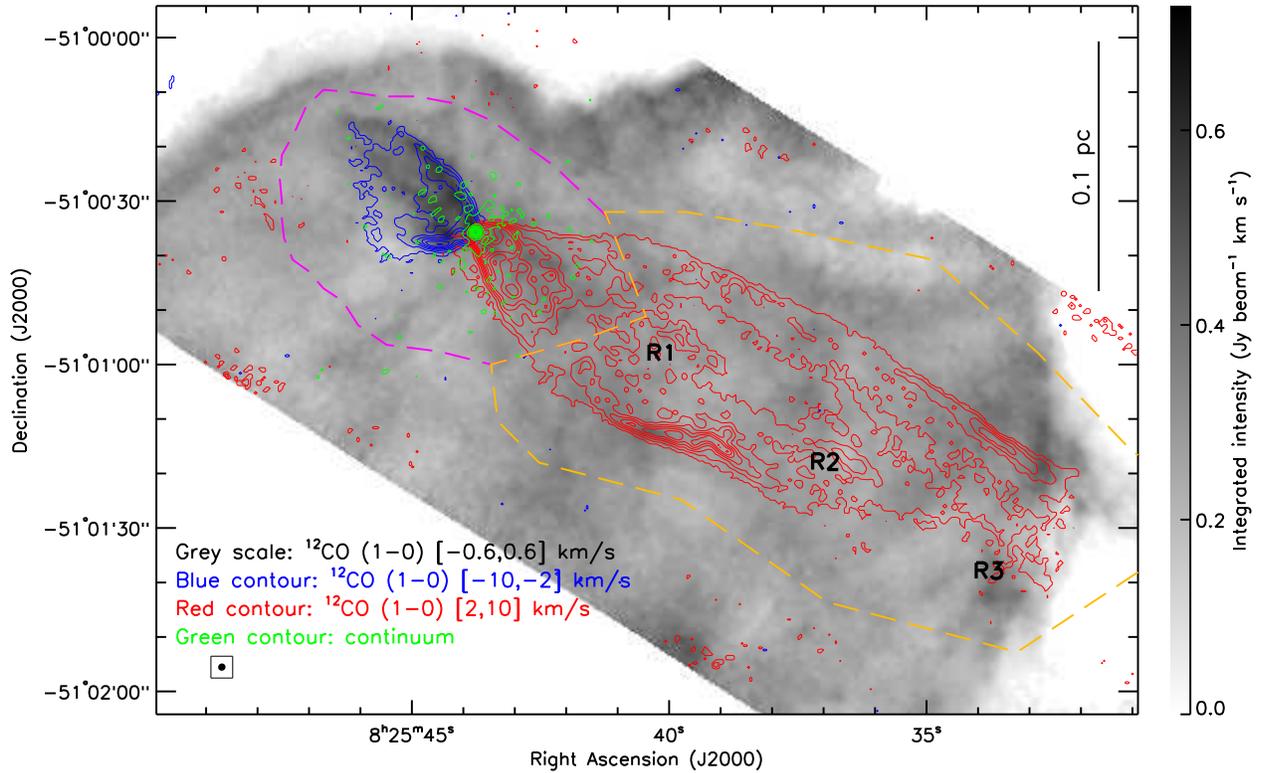}\\
\caption{Integrated intensity maps of the HH 46/47 $\twco$ (1-0) emission. 
The blue contours represent the blue-shifted emission integrated from $-10$ to $-2~\kms$ 
relative to the cloud velocity. The red contours represent the red-shifted emission
integrated from 2 to 10~$\kms$. The blue and red contours start at $3\sigma$ and have a step of $6\sigma$
($1\sigma=0.023~\Jybeamkms$). The grey scale represents the $\twco$ (1-0) emission integrated over 
velocity range from $-0.6$ to $0.6~\kms$ relative to the cloud velocity. 
The synthesized beam of $\twco$ is $1.37\arcsec \times 1.31\arcsec$ (P.A. = $-58.2^\circ$).
The green contours show the 100 GHz continuum emission. The contours start at 
$3\sigma$ and have a step of $15\sigma$ ($1\sigma=0.041~\mJybeam$).
The labels R1, R2, R3 mark the positions of the three bright clumps on the outflow axis 
(see Sections \ref{sec:12CO} and \ref{sec:outflow} for more details). 
The dashed lines define two sub-regions: the central region (purple lines) 
covering the the blue-shifted outflow and the base of the red-shifted outflow, 
and the extended reb lobe (yellow lines) covering the rest of the red-shifted outflow. 
The morphology of $\thco$ and $\ceio$ emission has been taken into account when defining these regions. 
These regions are used in Section \ref{sec:mass} in order to exclude emission which is not associated with the outflow.}
\label{fig:int_12CO}
\end{center}
\end{figure*}

\begin{figure*} 
\begin{center}
\includegraphics[width=\textwidth]{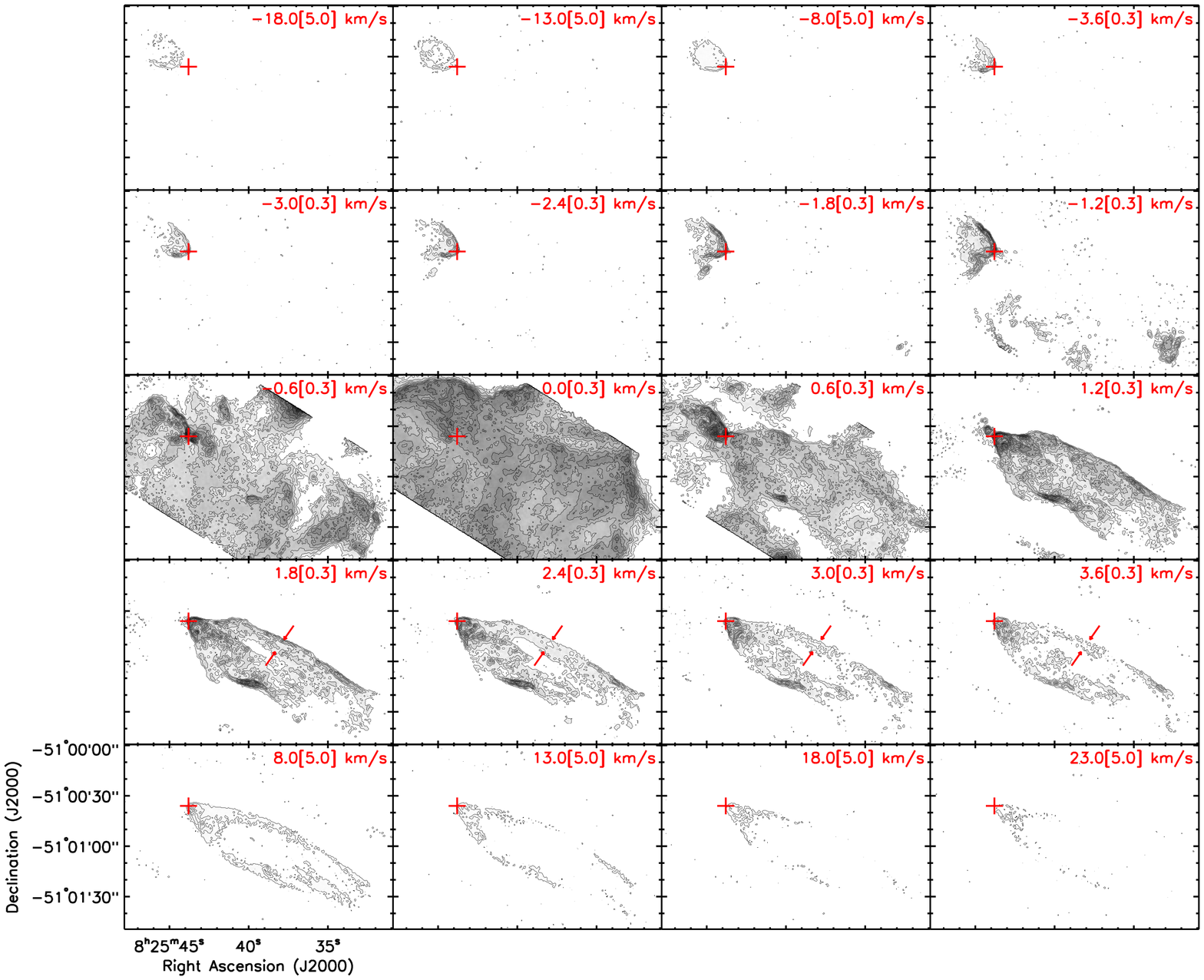}\\
\caption{Channel maps of the $\twco$ (1-0) emission. 
In the upper-right corner of each panel, the central outflow velocity relative to the cloud velocity
and the width of the channel (in brackets) are given.
The contours start at 3$\sigma$ with a step of 6$\sigma$.
1$\sigma$=3.4~$\mJybeam$ for channels with a width of 5~$\kms$, 
and 11~$\mJybeam$ for channels with a width of 0.3~$\kms$.
The synthesized beam is $1.37\arcsec \times 1.31\arcsec$ (P.A. = $-58.2^\circ$).
The red crosses mark the position of the peak of the continuum emission.
The red arrows in the panels on the fourth row mark where the thin structures on
the outflow cavity wall bifurcate (see Sections \ref{sec:12CO} and \ref{sec:outflow} for more details).}
\label{fig:chan_12CO}
\end{center}
\end{figure*}

The observations were carried out using ALMA from November 5, 2013 to April 11, 2015. 
Two different correlator configurations were used 
to provide images on six molecular lines and two 2 GHz-wide continuum bands in Band 3.
Data from both the 12m array and the ACA (including the 7m array and the 12m total power dishes) were obtained.

With the first correlator configuration, $\twco$ (1-0) at 115.27 GHz and 
$\cseo$ (1-0) at 112.36 GHz were observed along with continuum emission at 100.7 and 102.7 GHz (3 mm). 
The $\twco$ line was observed with a channel width of 61 kHz (0.2 $\kms$)
over a 117.2 MHz (305 $\kms$) bandwidth, and the C$^{17}$O line was observed 
with a channel width of 30.5 kHz (0.1 $\kms$)
over a 58.6 MHz bandwidth (152 $\kms$). 
The continuum emission was observed with two 1875 MHz-wide bands.
The 12m array data were obtained over 3 scheduling blocks, 
with 32 to 37 antennas, and projected baselines in the range of 12 to 528 m.
A rectangular 23-point mosaic with contiguous pointings separated by 25.8$\arcsec$ and oriented at a position
angle (P.A.) of about $60^\circ$ was used to map the outflow.
The 7m array data were obtained over 10 scheduling blocks,
with 8 to 10 antennas, and projected baselines in the range of 7 to 44 m. 
A 9-point mosaic with pointing separation of 44.2$\arcsec$ was used for mapping. 
The total power data were obtained over 18 scheduling blocks, with a mapping area of
$6\arcmin \times 3.8\arcmin$.
The resultant map combining the 12m array, 7m array and total power data has a size of about $2.1\arcmin \times 1.05\arcmin$ 
and is centered at $8\h25\m40\s$, $-51^\circ00\arcmin59\arcsec$ (J2000).
 
The second correlator configuration provides simultaneous observations of
$\thco$ (1-0) at 110.2 GHz, $\ceio$ (1-0) at 109.78 GHz, 
CS (2-1) at 97.98 GHz, and $\cths$ (2-1) at 96.41 GHz.
Each line was observed with a channel width of 30.5 kHz ($0.1~\kms$), 
and over a bandwidth of 58.6 MHz ($160~\kms$). 
The 12m array data were obtained over 3 scheduling blocks, with 29 antennas,
and projected baselines ranging from 12 to 340 m.
A 26-point mosaic with pointings separated by 27$\arcsec$ was used to cover a similar but
slightly wider area than the $\twco$ data.
The 7m array data were obtained over 10 scheduling blocks,
with 8 to 10 antennas, and projected baselines ranging from 7 to 44 m.
A 7-point mosaic with a pointing separation of 46.3$\arcsec$ was used. 
The total power data were obtained over 12 scheduling blocks, with an mapping area of
$6\arcmin \times 4.2\arcmin$.
The resultant combined map is about $2\arcmin \times 1.3\arcmin$ 
centered at $8\h25\m40\s$, $-51^\circ00\arcmin57\arcsec$ (J2000).

Ganymede, Pallas, J0538-440, Jupiter, Mars, Callisto, and J1256-0547 were used as gain and flux calibrators,
J0845-5448 and J0701-4634 were used as phase calibrators, and J0747-3310, J0922-3959, J0538-4405,
J1107-4449, J1037-2934 and J0519-4546 were used as bandpass calibrators. 
The data were edited, calibrated and imaged in CASA. The 12m array and the 7m array visibilities
were combined with their weighting factors estimated from the data noise using the CASA task {\it statwt}.
The combined interferometric data have projected baselines ranging from 7 to 525 m for $\twco$ and $\cseo$, 
and baselines ranging from 7 to 340 m for $\thco$, $\ceio$, CS and $\cths$. 
The data were imaged using the CLEAN algorithm. For the spectral data we defined
a different clean region for each channel, encircling the area with the brightest emission. 
Robust weighting with the robust parameter of 0.5 is used in the clean process.
The resulting synthesized beam is $1.3\arcsec \times 1.3\arcsec$ 
for the $\twco$ data cube and $3.2\arcsec \times 1.6\arcsec$ for the $\thco$ and $\ceio$ data cubes.

The interferometric data and the total power data were then combined in the image space using
the CASA task {\it feather}.
In the rest of the paper, if not indicated otherwise, we will show the combined data.
Figure \ref{fig:comb} shows the $\twco$ (1-0) integrated maps as an example of the combination. 
With a similar angular resolution as the 12m array data, 
the combined data keep the detailed structures revealed by the 12m array data,
while also showing the diffuse emission around the cloud velocity.
Details of the $\twco$ emission will be discussed in Section \ref{sec:12CO}.
Throughout the paper we define the outflow velocity $\vout$ as the LSR velocity of the emission minus
the cloud LSR velocity which is 5.3 $\kms$ (\citealt[]{vanKempen09}).

\section{Results}
\label{sec:results}

\subsection{Continuum}
\label{sec:contin}

Figure \ref{fig:continuummap} shows the continuum emission from the 12m array and 7m array
data with only the two 2GHz-wide spectral windows.
With the much higher sensitivity provided by the large bandwidth and the 
coverage of short baselines, our Cycle 1 data reveals a fainter extended structure in continuum 
in addition to the compact component.
This extended structure appears to be elongated
(about $10\arcsec \times 5\arcsec$, i.e. 4400 AU $\times$ 2200 AU) with 
its major axis perpendicular to the axis of the outflow.
This extended emission curves towards the direction of the red-shifted outflow.
In particular its southern part seems to follow the shape of the red-shifted outflow cavity. 
On the eastern side, the faintest emission also appears to follow the
shape of the blue-shifted outflow cavity. Therefore the extended continuum emission is likely tracing
a flattened envelope which is shaped by the outflow cavities on both sides.

The peak of the continuum emission is at $8\h25\m43\s.766$, $-51^\circ00\arcmin35\arcsec.70$ (J2000) 
and has an intensity of 6.4 $\mJybeam$ (0.33 K). The position is consistent with the ALMA Cycle 0 observation 
(Paper I) and previous HST observation of HH 47 IRS (\citealt[]{Reipurth00}).
The peak intensity is higher than previously observed (0.15 K from Paper I), which is likely
caused by the lower beam-dilution produced by the current beam 
(which is more than a factor of two smaller than that of the Cycle 0 observations). 
The angular resolution of our continuum observation is still not high enough to resolve the
possible different peaks associated with the binary 
(i.e. the individual circumstellar disks surrounding each of the binary companions)
which have a separation of 0.26$\arcsec$ (\citealt[]{Reipurth00}). 
The compact component seen here probably traces a circumbinary envelope.
Integrated over a circle with radius of 8$\arcsec$, 
the total flux density of the continuum emission is $11\pm 0.012$ mJy.
Such a total flux corresponds to a mass of 0.3 $M_\odot$, using the method described
in Paper I and assuming a dust temperature of 30~K, a gas-to-dust mass ratio of 100,
and an emissivity spectral index ($\beta$) of 1.
The dust opacity ($\kappa$) at 3 mm is estimated to be 0.9 $\mathrm{cm}^2~\mathrm{g}^{-1}$, by
extrapolating the value of $\kappa$ at 1.3 mm obtained by \citet[]{Ossenkopf94} for dust with a
thin ice mantle after $10^5$ yr of coagulation at a gas density of $10^6~\mathrm{cm}^{-3}$.

\subsection{$\twco$ (1-0)}
\label{sec:12CO}

The integrated emission of the $\twco$ (1-0) line is shown in Figure \ref{fig:int_12CO}.
Compared with the Cycle 0 observation, the resolution is improved
by a factor of 2.4, and the extended emission is recovered. 
The integrated map in the velocity range from $-0.6$ to $0.6~\kms$ relative to the cloud velocity
shows mainly the emission from the host globule with its edges highlighted.
The molecular outflow lobes stop at the edge of the globule,
therefore there is a drastic contrast in the sizes of the blue-shifted and the red-shifted
lobes.
The blue-shifted lobe is short and with most of its emission concentrated 
on the outflow cavity wall with a parabolic shape.
In contrast, being more deeply embedded, the red-shifted
outflow contains a large amount of gas inside of the outflow cavity.
Three bright clumps are seen inside of the red lobe along the outflow axis,
at distances of approximately $40\arcsec$, $80\arcsec$ and $110\arcsec$ 
from the central source (identified as R1, R2 and R3 in Paper I).
More extended emission is also detected connecting these clumps to the outflow cavity walls,
following the shapes of bow shocks with the apexes at the three clumps.
This is consistent with a scenario in which these structures 
trace material entrained by a series of bow-shocks caused by episodic ejection events in
the jet.
The blue-shifted outflow also shows a structure that appears at very low velocities (shown in grey scale),
tracing the northern cavity wall and a jet-like component 
(approximately along the outflow axis) connecting the base and the end
of the outflow lobe. It is unclear if the latter structure traces material entrained by
the blue-shifted jet seen at optical wavelengths.

\begin{figure*} 
\begin{center}
\includegraphics[width=\textwidth]{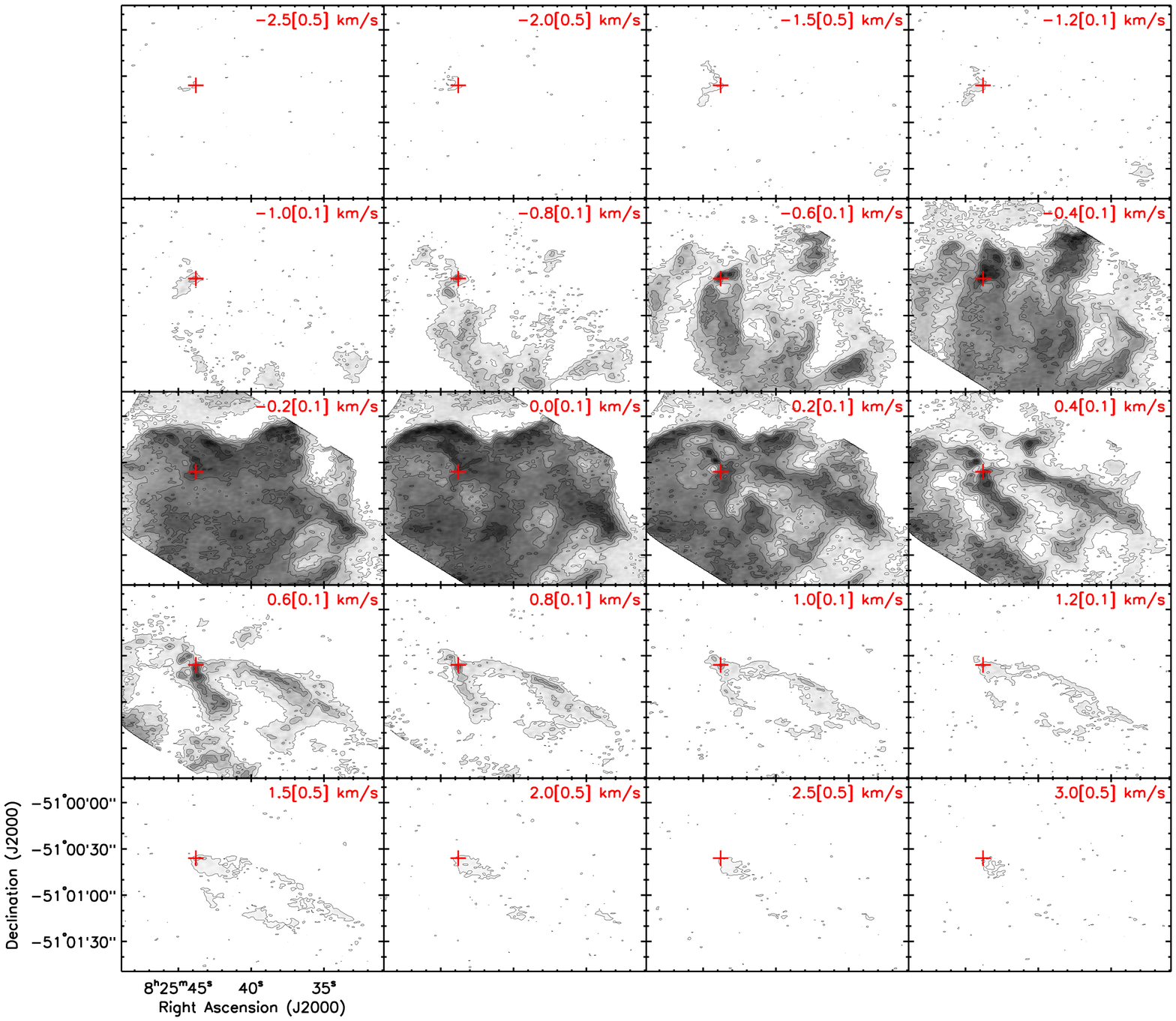}\\
\caption{Channel maps of the $\thco$ (1-0) emission. 
In the upper-right corner of each panel, the central outflow velocity relative to the cloud velocity
and the width of the channel (in brackets) are given.
The contours start at 3$\sigma$ with a step of 9$\sigma$.
1$\sigma$=7.4~$\mJybeam$ for channels with a width of 0.5~$\kms$, 
and 13~$\mJybeam$ for channels with a width of 0.1~$\kms$.
The synthesized beam is $3.18\arcsec \times 1.67\arcsec$ (P.A. = $-86.1^\circ$).
The red crosses mark the position of the peak of the continuum emission.}
\label{fig:chan_13CO}
\end{center}
\end{figure*}

\begin{figure*} 
\begin{center}
\includegraphics[width=0.85\textwidth]{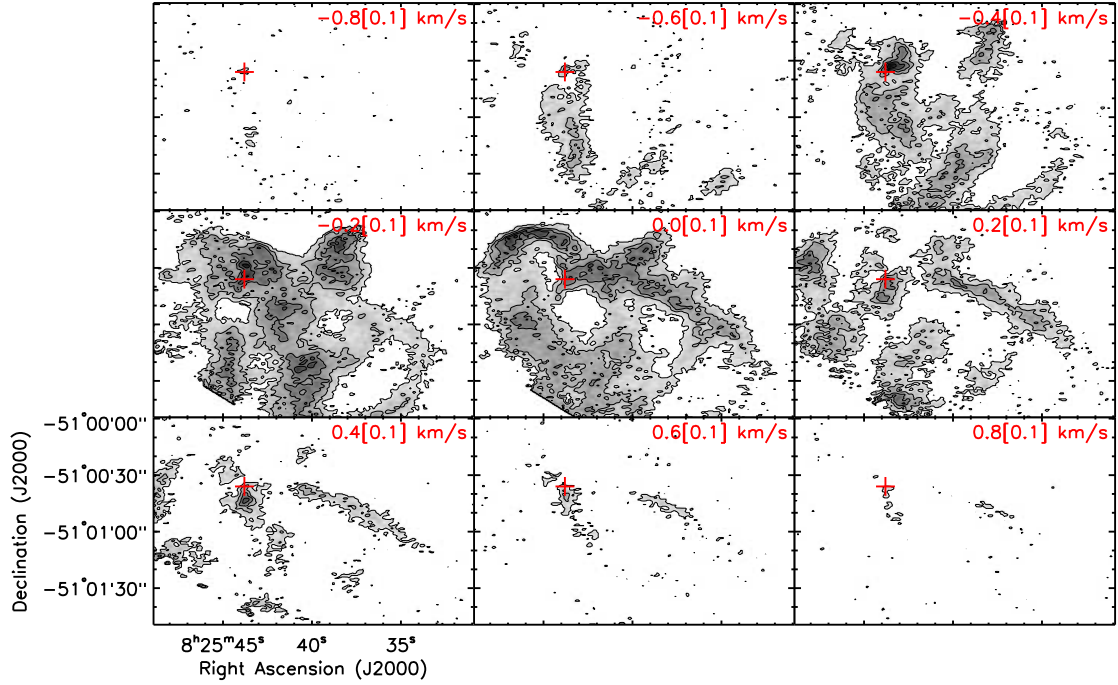}\\
\caption{Channel maps of the $\ceio$ (1-0) emission.
The central outflow velocity relative to the cloud velocity
and the width of the channel are shown in the upper-right corner of each panel.
The contours start at 3$\sigma$ with a step of 6$\sigma$ 
(1$\sigma$=13~$\mJybeam$).
The synthesized beam is $3.20\arcsec \times 1.69\arcsec$ (P.A. = $-86.0^\circ$).
The red cross shows the peak of the continuum emission.}
\label{fig:chan_C18O}
\end{center}
\end{figure*}

\begin{figure*} 
\begin{center}
\includegraphics[width=0.85\textwidth]{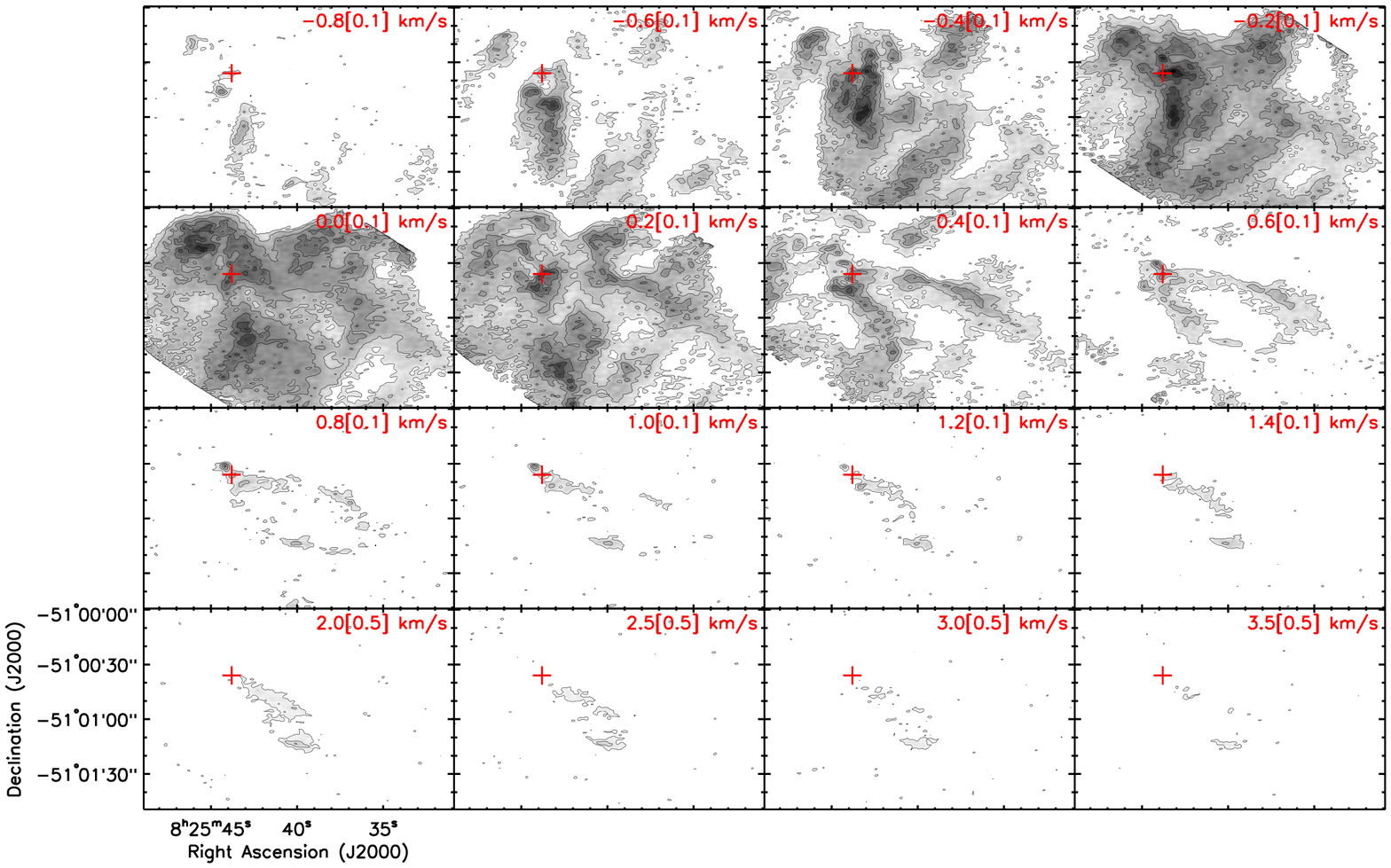}\\
\caption{Channel maps of the CS (2-1) emission.
The central outflow velocity relative to the cloud velocity
and the width of the channel are shown in the upper-right corner of each panel.
The contours start at 3$\sigma$ with a step of 6$\sigma$.
1$\sigma$=10~$\mJybeam$ for channels with a width of 0.1~$\kms$, and
5.5~$\mJybeam$ for channels with a width of 0.5~$\kms$. 
The synthesized beam is $3.62\arcsec \times 1.92\arcsec$ (P.A. = $-86.6^\circ$).
The red cross shows the peak of the continuum emission.}
\label{fig:chan_CS}
\end{center}
\end{figure*}

Figure \ref{fig:chan_12CO} shows the channel maps of the $\twco$ (1-0) emission,
with a channel width of 0.3 $\kms$ for the velocity range from $-3.6$ to $3.6~\kms$ relative
to the cloud velocity and a channel width of 5 $\kms$ for higher velocities.
We detect emission up to about $-30~\kms$ for the blue-shifted outflow and about $35~\kms$ for
the red-shifted outflow with the 5 $\kms$ channel width.
The blue-shifted outflow is seen at velocities $\le -0.9~\kms$.
At velocities $\gtrsim-3~\kms$, the blue-shifted outflow follows a parabolic shape
outlining the outflow cavity,
except there is a feature towards the southwest separated from
the main outflow. The latter is argued in Paper I to be a second outflow 
possibly driven by the binary companion of the protostar driving the main outflow.
However we are still not able to identify its counter-lobe.
At a higher velocity (e.g., $\le -8~\kms$), 
the emission forms elliptical rings and moves further away from the central source. 
This feature can be explained as outflowing shells entrained
by a wide-angle wind and will be discussed in detail in Section \ref{sec:outflow}.

The red-shifted outflow cavity structure is seen at velocities $\ge 1.2~\kms$.
As the velocity increases, the outflow cavity becomes narrower, which is most evident 
at the base of the outflow close to the central source.
The northern cavity wall of the red-shifted outflow shows at least two parallel thin structures 
at velocities between 1.8 and 3.6 $\kms$ (marked by red arrows in Figure \ref{fig:chan_12CO}).
Such structures were noticed in the Cycle 0 data (Paper I). 
With the new higher resolution data, we see that these layers start 
at only about $40\arcsec$ from the central source and extend to about $90\arcsec$ from the central source 
(i.e. nearly 0.13 pc on the plane of the sky).
The inner layer curves towards the R2 clump on the jet axis
(see the channel map at 2.4 $\kms$), while the outer layer extends further and curves towards 
clump R3 (e.g. the channel map at 3.6 $\kms$).
We believe these structures trace the material in shells formed in multiple jet bow-shock events
and we will discuss them further in Section \ref{sec:outflow}.

At velocities from $-0.6~\kms$ to $0.6~\kms$, most of the emission is associated with
the parent cloud. 
However at the velocity of $0.6~\kms$, bright emission is seen towards the northeast of the
central source, which overlaps with the HH 46 reflection nebula and the blue-shifted
outflow, and likely corresponds to the emission from the back side of the northeast (mostly blueshifted) lobe. 

\subsection{$\thco$ (1-0)}
\label{sec:13CO}

We show the channel maps of the $\thco$ (1-0) emission in Figure \ref{fig:chan_13CO},
with a channel width of 0.1 $\kms$ for velocities within 1.2~$\kms$ relative to the
cloud velocity, and a channel width of 0.5 $\kms$ for higher velocities.  
$\thco$ is a higher column density tracer than $\twco$, which allows it to
trace medium density (about $10^3~\cm^{-3}$) material in a globule (e.g. \citealt[]{Arce05}).
Here it is only detected at relatively low outflow velocities from $-2.5$ to $3.5~\kms$.
The blue-shifted outflow is seen at velocities from $-2.5$ to $-0.8~\kms$. 
At velocities from $-2.5$ to $-1.5~\kms$ the $\thco$ traces a V-shaped outflow cavity,
but with no emission inside the apparent cavity structure unlike $\twco$ at these velocities.
In addition the emission is brighter in the southern cavity wall as opposed to what is seen in $\twco$.
The second outflow towards the southeast of the central source is
seen at velocities from $-1.5$ to $-0.8~\kms$.
The emission at velocities from $-0.6$ to $0.2~\kms$ are dominated by the cloud material,
which shows clumpy structures towards the south and southwest of the central source. 
The red-shifted outflow cavity, especially the northern cavity wall, 
appears at a velocity as low as $0.2~\kms$. 
At velocities higher than 0.8 $\kms$ the $\thco$ emission mainly traces the limb-brightened
outflow cavity. The cavity forms a loop with its end coincident with
the $\twco$ clump R2. 
In the vicinity of the central source, the peak of the $\thco$ emission moves from north
of the source at velocities between approximately $-0.6$ and $0~\kms$ to south of the source at 
velocities between 0 and $0.6~\kms$.
The velocity gradient is perpendicular to the direction of the outflow and
suggests a rotating envelope, consistent with the gradient observed in $\ceio$ (see below).

\subsection{$\ceio$ (1-0)}
\label{sec:C18O}

We show the channel maps of the $\ceio$ (1-0) emission in Figure \ref{fig:chan_C18O}.
Being a higher column density tracer than $\thco$, $\ceio$ typically traces gas with densities 
approximately $10^4$ to $10^5~\cm^{-3}$ inside molecular clouds (e.g. \citealt[]{Fuller02}).
With a channel width of 0.1 $\kms$, the emission is only detected within 
1 $\kms$ from the cloud velocity.
At blue-shifted velocities, the emission traces the clumpy structures inside the
parent globule extending from the central source to the south and southwest,
which also appear in the $\thco$ (1-0) emission at these velocities.
The blue-shifted outflow cavity is not detected in this line.
The northern cavity wall of the red-shifted outflow starts to appear at a velocity
as low as 0.2 $\kms$ and dominates the morphology of the $\ceio$ emission
up to 0.8 $\kms$. The southern cavity wall of the red-shifted outflow appears
at velocities from 0.4 to 0.8 $\kms$ but only at regions close to the central
source. The limb-brightened outflow cavity in $\ceio$ may arise from material
accelerated by the outflow piled up along the cavity walls or from a higher
excitation temperature or a higher abundance due to
the interaction of the outflow and the core.
Only the red lobe is detected in this high column density tracer
because it is more embedded in the cloud than the blue lobe.
In the vicinity of the central source, the emission peak moves from
the north of the source at blue-shifted velocities (from $-0.6$ to $0~\kms$) to
the south of the source at red-shifted velocities (from 0 to $0.6~\kms$),
which could be interpreted as rotation (see Section \ref{sec:rotationcore}). 

\subsection{CS (2-1)}
\label{sec:CS}

In Figure \ref{fig:chan_CS} we show the channel maps of the CS (2-1) emission.
CS is typically a tracer of material at densities greater than approximately 
$10^4~\cm^{-3}$ (\citealt[]{Mardones97}). 
Between about $-0.8$ and 0.2 $\kms$, most of the emission traces the cloud structures
towards the south and southwest of the central source, which also appear in $\thco$
and $\ceio$.
The blue-shifted outflow cavity is not detected, while the red-shifted outflow cavity 
appears at velocities from about $0.2$ to $1~\kms$.
Especially around 0.6 to $0.8~\kms$ the emission forms a loop outlining
the red-shifted outflow cavity with its end coinciding with the $\twco$
clump R2. 
Above $1~\kms$ the CS emission mainly traces a collimated structure inside the outflow cavity.
The region where the southern outflow cavity wall ends is also visible in CS at these velocities.
The jet-like structure in CS
overlaps with the $\twco$ emission inside the outflow cavity along the length of the infrared jet
(as seen in the maps presented by \citealt[]{Noriega04} and \citealt[]{Velusamy07}),
suggesting that the CS also probes the jet-entrained material inside
the outflow cavity. 
The kinematics of this structure will be discussed in Section \ref{sec:rotationoutflow}.

\subsection{$\cseo$ (1-0) and $\cths$ (2-1)}
\label{sec:C17O_C34S}

\begin{figure} 
\begin{center}
\includegraphics[width=\columnwidth]{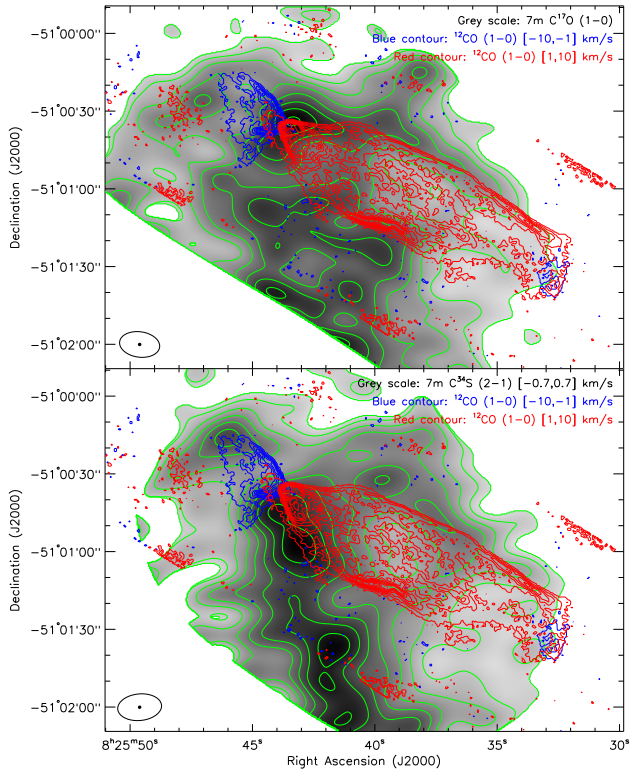}\\
\caption{{\bf Upper panel:} The integrated emission of the $\cseo$ (1-0) line combining the 7m array and
total power data in grey scale and green contours, overlaid with the blue contours and red contours 
showing the $\twco$ (1-0) emission integrated in the velocity range from $-10$ to  $-1~\kms$ and from 1 to $10~\kms$. 
The $\cseo$ (1-0) emission is integrated over 5 $\kms$ to cover the three
hyperfine lines spanning 1.2 MHz (about $4~\kms$). The green contours start at 
$3\sigma$ and have a step of $3\sigma$ ($1\sigma=0.043~\Jybeamkms$)
The blue and red contours start at $3\sigma$ and have a step of $6\sigma$
($1\sigma=0.023~\Jybeamkms$). 
The synthesized beam of $\cseo$ is $15.6\arcsec \times 9.8\arcsec$ (P.A. = $-82.1^\circ$).
{\bf Lower panel:} Same as the upper panel but showing the $\cths$ (2-1) emission combining the 7m array 
and the total power data integrated from $-0.7$ to $0.7~\kms$ in grey scale and green contours. 
The green contours start at $3\sigma$ and have a step of $3\sigma$ ($1\sigma=0.016~\Jybeamkms$)
The synthesized beam of $\cths$ is $17.0\arcsec \times 10.3\arcsec$ (P.A. = $-84.6^\circ$).}
\label{fig:int_C17O_C34S}
\end{center}
\end{figure}

The $\cseo$ (1-0) emission of the HH 46/47 outflow is only detected in the 7m array and total power data,
which is shown in the upper panel of Figure \ref{fig:int_C17O_C34S}.
We detect the three hyperfine lines of $\cseo$ (1-0), with a width of about 0.5 $\kms$ for each of them.
Towards the central source there is bright $\cseo$ emission with its peak slightly to the northwest
of the central source, tracing the immediate envelope.
Extending to the west, $\cseo$ appears to follow the shape of the red-shifted $\twco$ outflow,
with little emission inside of the cavity, 
clearly suggesting that the outflow has created a low column density cavity from the parent globule.
There is a distinct $\cseo$ clump close to the bright $\twco$ emission 
on the southern wall of the red lobe, which we believe is a dense clump that the outflow is interacting with
and produces the kink in the $\twco$ cavity wall. 
The high column density material also extends further to the south of the central source, similar to
the $\thco$ and $\ceio$ maps.

The $\cths$ (2-1) emission is also only detected in the 7m array and total power data
within the velocity range from $-0.6$ to 0.6~$\kms$,
which is shown in the lower panel of Figure \ref{fig:int_C17O_C34S}.
The $\cths$ appears to trace a linear structure inside the globule 
starting from where the central source is and extending to the south.
The cavity carved out by the red-shifted outflow is also seen. 

\subsection{Mass, Momentum, and Kinetic Energy of the Outflow}
\label{sec:mass}

To accurately estimate the mass, momentum and energy of the CO outflow
requires properly correcting for the optical depth of the lines used for estimating the column densities. 
The fact that most of the outflowing
material is at low velocities and that $\twco$ is optically thick at these velocities
lead to a severe underestimate of the mass if simply assuming it is optically thin
(e.g. \citealt[]{Arce01}; \citealt[]{Dunham14}). One way to estimate the optical depth of $\twco$ is to
use the intensity ratio between $\twco$ and one of its optically thinner isotopologues (usually
$\thco$). However $\thco$ itself maybe optically thick at low velocities ($\lesssim 1~\kms$).
Here, with observations of $\twco$, $\thco$, and $\ceio$ with similar sensitivities
and angular resolutions we are able to estimate the $\twco$ and $\thco$ optical depths 
and therefore much more accurately measure the outflow properties.

We estimate the correction factor to the optical depth following the method outlined
by \citet[]{Dunham14} but applied to three CO isotopologues $\twco$, $\thco$ and $\ceio$. 
A detailed formulation can be found in Appendix \ref{app:mass}.
Assuming $\twco$, $\thco$, and $\ceio$ approximately trace the same material 
and have the same excitation temperature
$\Tex$ and beam filling factor, and that $\ceio$ is optically thin, we have
\begin{eqnarray}
\frac{T_{R,13}(v)}{T_{R,18}(v)} & = &\frac{1-\exp(-\tau_{v,13})}{1-\exp(-\tau_{v,18})}\nonumber
=\frac{1-\exp(-\tau_{v,13})}{\tau_{v,18}}\\
&\approx& X_{13,18}\frac{1-\exp(-\tau_{v,13})}{\tau_{v,13}},
\end{eqnarray}
where the subscripts 13 and 18 represent $\thco$ and $\ceio$ respectively,
and $X_{13,18}$ is the abundance ratio between $\thco$ and $\ceio$. 
To correct for the optical depth of $\thco$, we simply need to multiply the measured $\thco$ intensity by 
a factor
\begin{equation}
F_{\tau,13}(v)\equiv \frac{\tau_{v,13}}{1-\exp(-\tau_{v,13})}=X_{13,18}\frac{T_{R,18}(v)}{T_{R,13}(v)}.
\end{equation}
Similarly, the optical depth correction factor for $\twco$ is
\begin{equation}
F_{\tau,12}(v)=X_{12,13}\frac{T'_{R,13}(v)}{T_{R,12}(v)},
\end{equation}
where $T'_R\equiv T_{R}(v)F_{\tau}$ is the optical-depth corrected intensity.
Note, the intensity ratio has an upper limit equal to the abundance ratio, therefore the correction
factors have a lower limit of 1.
In this paper, we adopt $X_{12,13}=62$ (\citealt[]{Langer93}), $X_{13,18}=8.7$ 
(from [$^{16}$O]/[$^{18}$O]=540, \citealt[]{Wilson92}).

\begin{figure} 
\begin{center}
\includegraphics[width=\columnwidth]{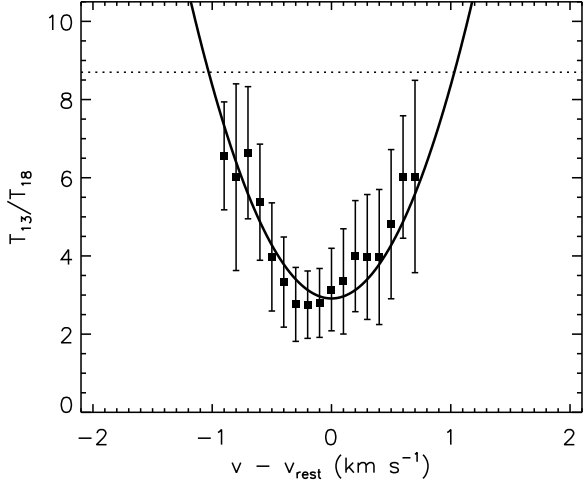}\\
\caption{Intensity ratio between $\thco$ (1-0) and $\ceio$ (1-0) as a function of velocity. 
At each velocity, the data point and the error bar are the weighted mean and weighted standard deviation
of the intensity ratios over the pixels where both $\thco$ (1-0) and $\ceio$ (1-0) are detected above 5$\sigma$
(see text). Only the pixels within the two defined sub-regions 
(see Figure \ref{fig:int_12CO}) are included.
The solid curve is the best-fit second-order polynomial, 
and the dotted lines indicate the abundance ratio which sets the upper limit of the intensity ratio.}
\label{fig:opac_13CO}
\end{center}
\end{figure}

Figure \ref{fig:opac_13CO} shows the mean intensity ratios between $\thco$ (1-0) and $\ceio$ (1-0)
as a function of velocity. 
In each velocity channel, we first calculate the intensity ratios between the two lines 
at the pixels where both lines are detected above 5$\sigma$. Only the pixels within the defined
sub-regions shown in Figure \ref{fig:int_12CO} are included since in this work we only focus on
the outflow.
The uncertainties of these pixel intensity ratios are calculated from the channel map rms errors
using error propagation.
We then calculate the weighted mean and weighted standard deviation
of these pixel intensity ratios in each channel (data points and error bars in the figure). 
The weights are inversely proportional to the
square of the uncertainties of the pixel intensity ratios. 
In such a way, the pixels with both lines detected with higher S/N carry more weight.
In order to estimate the optical depth of $\thco$
at higher velocities where the $\ceio$ line is not detected, we fit a parabola to the measured intensity ratios
with the minimum point fixed at zero velocity (\citealt[]{Arce01}; \citealt[]{Dunham14}; \citealt[]{Offner11}). 
Note that the intensity ratio reaches its upper limit, the abundance ratio between the two molecules, 
at velocities where both transitions are optically thin.
The fitted parabola (solid curve) is
\begin{equation}
T_{13}/T_{18}=(2.91\pm 0.40)+(5.46\pm1.48)(v-v_{\mathrm{rest}})^2
\end{equation}
with a reduced $\chi^2$ of 0.22.
According to this, the $\thco$ (1-0) line becomes optically thin at velocities higher than $1~\kms$. 

\begin{figure} 
\begin{center}
\includegraphics[width=\columnwidth]{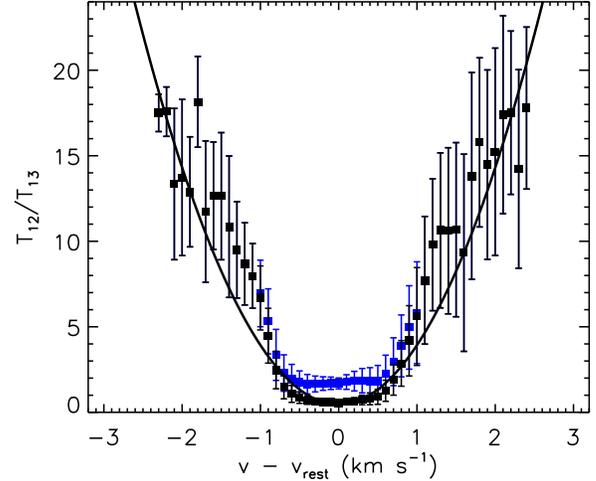}\\
\caption{Same as Figure \ref{fig:opac_13CO}, but showing the intensity ratios
between $\twco$ (1-0) and the optical-depth-corrected $\thco$ (1-0) emission
(black symbols, error bars). 
The blue symbols and error bars are the ratios and uncertainties using
the original $\thco$ intensities. 
The solid curve is the best-fit second-order polynomial using the black data points.}
\label{fig:opac_12CO}
\end{center}
\end{figure}

Figure \ref{fig:opac_12CO} shows the mean intensity ratios between $\twco$ (1-0) and $\thco$ (1-0)
as a function of velocity. 
The ratios are calculated in the same way as the $\thco$-to-$\ceio$ intensity ratios.
Again only the emission within the outflow region is included in the calculation.
Without correcting the optical depth of $\thco$, the intensity ratios become flat at
low velocities (blue symbols), suggesting optically thick $\thco$ emission. 
After correcting for the $\thco$ optical depth, we can fit a parabola even at the low velocities. 
The best fit for the region of the whole outflow is
\begin{equation}
T_{12}/T'_{13}=(0.47\pm0.053)+(3.46\pm0.14)(v-v_{\mathrm{rest}})^2
\end{equation}
and the reduced $\chi^2$ is 0.82.
From this fit, $\twco$ (1-0) becomes optically thin at velocities higher than 4.2 $\kms$.

To calculate the mass of the molecular outflow requires an estimate of the CO excitation temperature ($\Tex$).
There are various estimates of $\Tex$ for this outflow in the literature.
\citet[]{Chernin91} estimated an outflow excitation temperature of $8.4\pm1$ K using the 
$\twco$ (3-2) and $\twco$ (2-1) brightness temperature ratio. \citet[]{Olberg92} estimated
an excitation temperature of 15 K at outflow velocities using the intensity ratio between $\twco$ (1-0)
and $\twco$ (2-1). \citet[]{vanKempen09} estimated an excitation temperature of about 100 K along
the red-shifted outflow axis, and about 60 K for the blue-shifted outflow, using low and high-J transitions of CO.
We also can estimate the excitation temperature from the measured optically 
thick brightness temperature of $\twco$ at low velocities.
Within 4 $\kms$ relative to the cloud velocity where the $\twco$ (1-0) is optically thick,
from Eq. \ref{eq:rt} we can estimate the excitation temperature to be
$\Tex=5.53/\ln[1+5.53/(T_R+0.82)]$ (assuming a beam filling factor of 1). 
At these velocities, the peak intensities of the $\twco$ (1-0) line range from about 13 to 45 K, 
i.e., $\Tex=16 - 49$~K, consistent with the excitation
temperature from previous observations and appropriate for the gas traced by CO (1-0) line. 
In the rest of the paper, we will calculate
the mass and other properties of the outflow using two typical values of $\Tex$, 15 K and 50 K.

To calculate the mass and other properties of the outflow also requires separating the outflow
material from the cloud material. At low velocities, e.g. less than 1 $\kms$ relative to the cloud
velocity, the outflow cavity structure is already apparent, but there is also 
considerable emission from cloud material (not associated with the outflow).
One way to disentangle these two components is to fit the low velocity part
of the mass spectrum with a Gaussian distribution, and subtract such a component from the total
emission (e.g. \citealt[]{Arce01}; \citealt[]{Dunham14}). 
However, instead of just a total value, we are interested in the spatial distribution of the
outflow material. Therefore we simply apply a velocity boundary to exclude the contribution of the cloud material. 
For $\twco$, we only include the emission with $|\vout|\ge0.9~\kms$ in the central region and
$\vout \ge 0.9~\kms$ or $\vout \le -1.2~\kms$ in the extended red lobe. For $\thco$, we only include the emission
with $\vout\le-0.8~\kms$ or $\vout\ge0.4~\kms$ in the central region and $\vout\le-0.9~\kms$ or $\vout\ge0.4~\kms$ 
in the extended red lobe.
For $\ceio$, we only include the emission with $\vout\ge0.2~\kms$ in the extended red lobe.
We then calculate the column density of $\twco$, $\thco$ and $\ceio$ 
(after correcting for the optical depth) at each velocity channel, 
using Equation \ref{eq:dndv} assuming LTE conditions and a beam filling factor of 1.
We adopt an abundance of $\twco$ of $10^{-4}$ relative to $\Htwo$, and a gas mass of 
$2.34 \times 10^{-24}$ g per $\Htwo$ molecules.
Combining the column densities calculated from the three CO isotopologues in different pixels and velocity channels,
we obtain a combined column density map.
At the pixels where more than one CO isotopologues are detected, we choose the highest column density 
calculated from these CO isotopologues.

\begin{figure} 
\begin{center}
\includegraphics[width=\columnwidth]{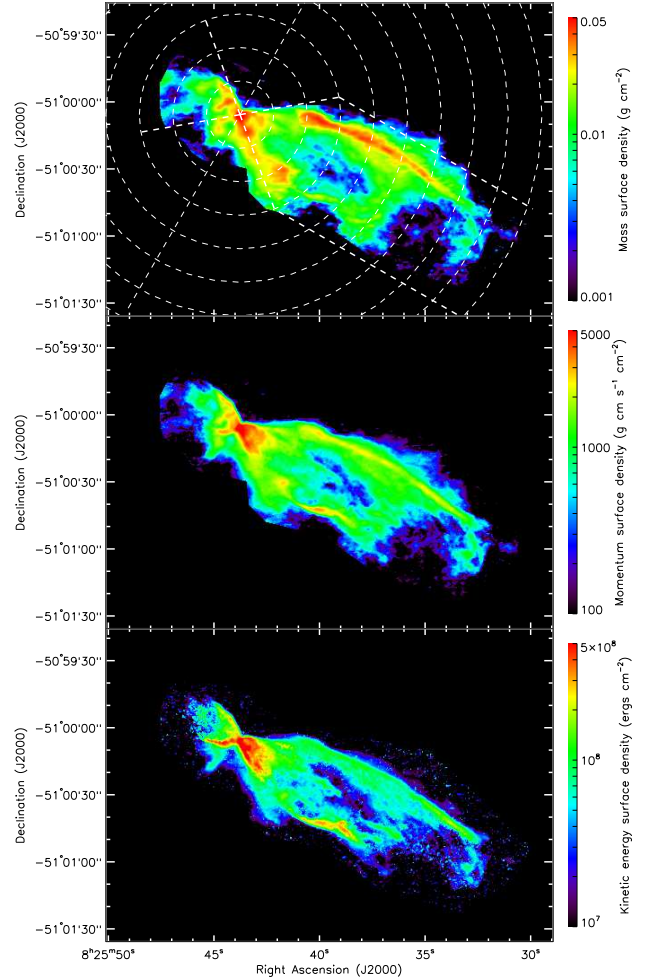}\\
\caption{Surface density maps of the mass, momentum and kinetic energy (from top to bottom) of the HH 46/47 
outflow. Only the emission considered to be associated with the outflow has been taken into account. 
The maps combine the material traced by $\twco$, $\thco$, and $\ceio$ (see text).
The dashed circles show the annuli over which we integrate to obtain the mass distribution with respect
to the distance from the central source in Figure \ref{fig:massdis}. The dashed diagonal line divides the
blue lobe and the red lobe. The other dashed lines roughly outline the outflow cavities which are used to 
estimate the mass of the ambient material which originally filled the outflow cavity (see Figure \ref{fig:massdis}
and Section \ref{sec:compare}).}
\label{fig:massmap}
\end{center}
\end{figure}

\begin{figure} 
\begin{center}
\includegraphics[width=\columnwidth]{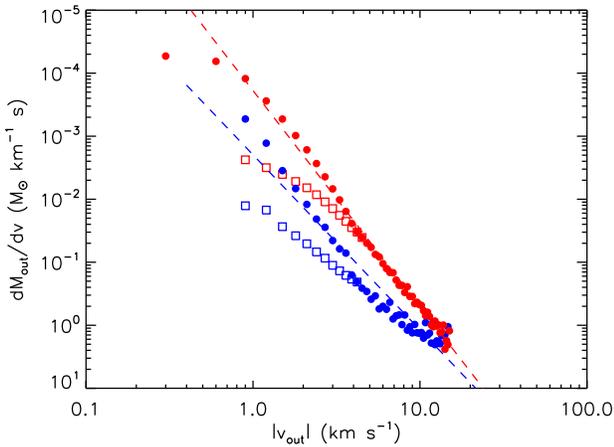}\\
\caption{The mass spectra of the HH 46/47 outflow. The blue and red symbols are for 
the blue-shifted and red-shifted material respectively.
The open squares show the mass derived from $\twco$ without optical depth correction. 
The filled circles show the mass obtained combining the optical-depth corrected 
$\twco$ and $\thco$ emission, and the $\ceio$ data.
The dashed lines are power-law fits to the filled circles within the velocity range 
from 0.6~$\kms$ (red-shifted) or 0.9~$\kms$ (blue-shifted)
to 15~$\kms$.}
\label{fig:massspec_out1}
\end{center}
\end{figure}

In Figure \ref{fig:massmap} we show the spatial distributions of the mass, momentum and
kinetic energy of the outflow, combining the three CO isotopologues and using
$\Tex=15$~K as an example.
The momentum surface density is defined as $\Pout=\Sigma_{\vout} M (x,y,\vout) \vout$
and the energy surface density $\Eout=\Sigma_{\vout} M (x,y,\vout) \vout^2/2$,
where $M (x,y,\vout)$ is the mass surface density of the outflow in each velocity channel,
and $\vout$ is not corrected for inclination.
Therefore they only provide lower limits for the real momentum and energy distributions.
Compared with the mass map, the momentum and energy 
is more concentrated towards the outflow cavity walls around the central source,
as they are more dominated by the high velocity material. 
In the kinetic energy distribution map (lower panel of Figure \ref{fig:massmap}), 
we see highlighted outflow cavity walls 
and even jet-like structures at the base of the outflow on both side.
The southern wall of the blue lobe is more prominent than the northern wall in the 
energy map, in contrast to the mass map.
The three shocked regions inside of the red lobe (R1, R2 and R3) are clearly seen in the
energy map, however, most of the energy is not around these clumps which mark the apexes 
of a series of jet bow-shocks, as it is predicted by jet-entrainment models,
rather it is along the cavity walls at the base of the outflow.
This suggests that even though the jet bow-shock features dominate the morphology
of the red lobe (see discussions in Section \ref{sec:outflow}), the contribution of a wide-angle
wind may dominate the energy input.
This is also supported by the fact that the blue-shifted and the red-shifted outflows 
appear to be more symmetric in the energy map (the brightest part),
since the high-velocity component of the blue-shifted outflow is mainly entrained by 
a wide-angle wind (see Section \ref{sec:outflow}).

\begin{table*} 
\begin{center}
\caption{Mass, momentum and kinetic energy of the outflow \label{tab:outflow}}
\ \\
\begin{tabular}{ll|cc|cc|cc}
\hline
\multirow{2}{*}{Tracer} & \multirow{2}{*}{Lobe$^\mathrm{a}$} & \multicolumn{2}{|c}{Mass$^\mathrm{b}$ ($10^{-2}~M_\odot$)} 
& \multicolumn{2}{|c}{Momentum$^\mathrm{b,c}$ ($10^{-2}~M_\odot~\kms$)} 
& \multicolumn{2}{|c}{Kinetic Energy$^\mathrm{b,c}$ ($10^{42}$ erg)} \\
\cline{3-8}
 & & $\Tex=15$~K & 50~K & 15~K & 50~K & 15~K & 50~K \\
\hline
\multirow{2}{*}{$\twco$} & Blue & 8.7 (1.0) & 22 (2.4) & 11/18 (2.5/3.8) & 28/44 (6.1/9.5) 
&  2.4/5.7 (1.2/3.0) & 5.8/14 (3.0/7.4) \\
& Red  & 49 (5.8) & 120 (14) & 64/99 (13/20) & 157/244 (32/50) & 11/28 (4.6/11) & 28/68 (11/27) \\
\hline
\multirow{2}{*}{$\thco$} & Blue & 3.5 (3.0) & 8.8 (7.6) &  3.6/5.6 (3.2/5.0) & 9.1/14 (8.1/13) 
& 0.43/1.0 (0.39/0.95) & 1.1/2.6(0.98/2.4) \\
& Red  & 98 (52) & 244 (129) & 53/83 (31/48) & 133/207 (78/121) & 3.4/8.3 (2.3/5.6) & 8.6/21 (5.8/14) \\
\hline
\multirow{2}{*}{$\ceio$} & Blue & 0 & 0 & 0/0 & 0/0 & 0/0 & 0/0 \\
& Red  & 37 & 93 & 12/19 & 30/46 & 0.46/1.1 & 1.2/2.8 \\
\hline
\multirow{2}{*}{Combined$^\mathrm{d}$} & Blue & 10 & 25 & 13/20 & 32/50 & 2.7/6.2 & 6.3/15 \\
& Red  & 152 & 378 & 110/170 & 271/422 & 14/33 & 34/82 \\
\hline
\end{tabular}
\end{center}
Notes:\\
$^\mathrm{a}$ Blue (red) indicate all outflow emission at blueshifted (redshifted) velocities 
with respect to the cloud velocity, independent of position.\\ 
$^\mathrm{b}$ The values outside of the parenthesis are with optical depth corrections, 
and those in parenthesis are without such correction.\\
$^\mathrm{c}$ The values before the slash are not corrected for the outflow inclination and those after
the slash are corrected assuming an inclination of 40$^\circ$ between the outflow axis and the plane of the sky.\\
$^\mathrm{d}$ Combining the $\twco$, $\thco$, and $\ceio$ emission in different positions and velocity channels 
(see the main text for more detail).
\end{table*}

In Figure \ref{fig:massspec_out1} we show the velocity distribution of the outflow mass,
assuming an excitation temperature of 15 K.
The correction for the CO optical depth and including the higher column density tracers
significantly increases the estimated mass at velocities $\lesssim 4~\kms$ 
(by more than an order of magnitude at velocities below $1~\kms$).
The slopes of the mass spectra become much steeper after such corrections.
We fit the combined mass spectra with power laws in a form of
$m(v)\propto v^{-\gamma}$, in the velocity range of $0.6\le|\vout|\le15~\kms$ for the red-shifted outflow and
$0.9\le|\vout|\le15~\kms$ for the blue-shifted outflow.
At lower velocities, the mass spectrum becomes flatter, which is due to the fact 
that in these channels the outflow structure can only be clearly identified in $\ceio$ and $\thco$ emission,
and the $\twco$ emission is excluded due to the cloud emission. 
While taking this low-velocity mass into account when calculating the total mass, 
we exclude it in fitting the slope of the mass-velocity relation.
We found a slope of $\gamma=-3.43\pm0.04$ for the red-shifted outflow and a slope of $\gamma=-2.78\pm0.11$ 
for the blue-shifted outflow. 
The red-shifted outflow is much more massive than the blue-shifted outflow,
but they have similar masses at velocities higher than 10 $\kms$.
\citet[]{Smith97} suggested that the mass spectrum slope steepens as the outflow
evolves, since as time goes by the material that once had been accelerated starts to slow
down gradually. In such a scenario, an outflow in a denser medium will
decelerate faster than an outflow in a low density medium (\citealt[]{Arce01}).
This may be an explanation for the steeper mass spectrum in the red-shifted outflow than
the blue-shifted outflow in this source.
Previous observations also suggested that $\gamma$ changes at a velocity
of about $10~\kms$ with a steeper power law for the higher velocity (e.g. \citealt[]{Richer00}).
We do not see such two distinct components in this source. Actually, the slope we find here is much steeper
than the slopes previously reported for the low velocity range $v<10~\kms$ ($-2.5<\gamma<-1$), but consistent with
the slopes reported for the velocity greater than $10~\kms$ ($-4<\gamma<-2.5$), which implies that
the previously reported change of slope at about $10~\kms$ may be due to uncorrected CO opacity.
However, we note that, even with optical depth correction, a few outflows still show change of slope 
in the mass spectrum around $10~\kms$ (e.g. \citealt[]{Su04}). 

Table \ref{tab:outflow} lists the total masses $\Mout=\Sigma_{x,y,\vout}M(x,y,\vout)$, 
the momentum $\Pout=\Sigma_{x,y,\vout} M (x,y,\vout) \vout$,
and the kinetic energy $\Eout=\Sigma_{x,y,\vout} M (x,y,\vout) \vout^2/2$
of the red-shifted and blue-shifted outflows, 
measured from $\twco$, $\thco$, $\ceio$ emission and combined,
both with and without optical depth corrections.
For the momentum and  kinetic energy, we also list the values after correcting for the inclination of the outflow
with respect to the plane of the sky $i$, which is assumed to be $40^\circ$ (see Section \ref{sec:outflow}).
The correction factor is $1/\sin i$ for $\Pout$ and $1/\sin^2 i$ for $\Eout$.
Note that these correction factors are only valid for outflows where all the motion is along the axis.
\citet[]{Downes07} constructed models of jet-entrained outflow and investigated the effect of inclination correction
on momentum and energy estimates, taking into account the transverse motions of the outflow. 
They found that the correction factor of $1/\sin i$ for $\Pout$ 
always overestimates the true momentum while the uncorrected $\Pout$ actually agrees with the true value.
They also found that $\Eout$ with the $1/\sin^2 i$ correction overestimates
the true value while the value without the inclination correction underestimates the true energy.

Using only $\twco$ and without correcting for the optical depth, our estimated 
outflow properties (the values in brackets in the first two rows of Table \ref{tab:outflow}) 
are similar to those measured in Paper I with ALMA cycle 0 data.
The correction of optical depth increases the mass estimation by a factor of about 8.5 for $\twco$.
This increase is consistent with previous observations of other
outflows (e.g. \citealt[]{Dunham14}) and simulations (e.g. \citealt[]{Offner11}; \citealt[]{Bradshaw15}).
The total momentum is increased by a factor of 4.9, and the total kinetic energy by a factor of
2.4 after the optical depth correction for $\twco$, suggesting that the momentum and kinetic
energy is dominated by material at velocities higher than $4~\kms$, where the $\twco$ line
becomes optically thin.
Completing the $\twco$ column density map with the slower material traced only by $\thco$ and $\ceio$, 
the total mass of the red-shifted outflow (last row of Table \ref{tab:outflow}) 
is 3 times of what is estimated using only opacity-corrected $\twco$, 
but the blue-shifted outflow mass is similar. The combined
total momentum is about $60\%$ higher than that measured using only $\twco$. The kinetic energy
estimated from the combined map is similar to that estimated from only $\twco$.
Therefore, only using the opacity-corrected $\twco$ emission may still underestimate the mass of the outflow
by a factor of 3 due to the lack of ability of $\twco$ to trace the low-velocity components,
but should represent a good estimate for the total momentum and the total kinetic energy.

With $\Tex=15$ K, the measured total mass of the CO outflow is 1.6 $M_\odot$, the total momentum is 
1.9 $M_\odot~\kms$ (after correcting for inclination) and the total energy is $3.9\times 10^{43}$ erg (after
correcting for inclination). The estimated mass and momentum of the CO outflow are significantly higher
than those estimated from surveys, which give average outflow masses of 0.09 $M_\odot$ for Class 0 sources 
and 0.06 $M_\odot$ for Class I sources, and average outflow momenta of 0.7 and 0.3 $M_\odot~\kms$ for 
Class 0 and I sources respectively (e.g. \citealt[]{Curtis10}). Similar low values were also given by \citet[]{Arce06}. 
With $\thco$ used to correct for the $\twco$ optical depth, \citet[]{Dunham14} estimated
the masses of 17 outflows from 0.01 to 0.8 $M_\odot$ and their momenta from 0.02 to 3 $M_\odot~\kms$. 
Our results are consistent with the highest mass and momentum they found, but higher than most of their sources. 
The main factor contributing to the differences between our results and previous estimates
is including higher column density tracers to trace outflowing gas at low velocities (around $1~\kms$).
Our estimate of the outflow energy is similar to the previously estimated values since the energy
is dominated by less dense, high velocity material traced by $\twco$.
Note that our estimates of the outflow mass, momentum and energy would be increased by a factor
of 2.5 with the higher $\Tex=50$ K.

The measured mass of the red-shifted outflow is 15 times higher than the mass of the blue-shifted outflow.
This again agrees with the very different environment of the two lobes.
The momentum and kinetic energy of the red lobe are still higher than those of the blue lobe, but by a 
smaller factor,
suggesting that even though much more material is entrained in the red lobe, the amount of material
at high velocities is similar on both sides, which is also seen in  Figures \ref{fig:massmap} 
and \ref{fig:massspec_out1}. 
We also see that the contrast between the two lobes is higher with a tracer of denser material (e.g., the contrast
is higher in $\thco$ than in $\twco$), suggesting there is more slower and denser material in the red lobe
than in the blue lobe.

\section{Discussion}
\label{sec:discuss}

\subsection{Role of the Outflow in Dispersing the Core}
\label{sec:compare}

\begin{figure} 
\begin{center}
\includegraphics[width=\columnwidth]{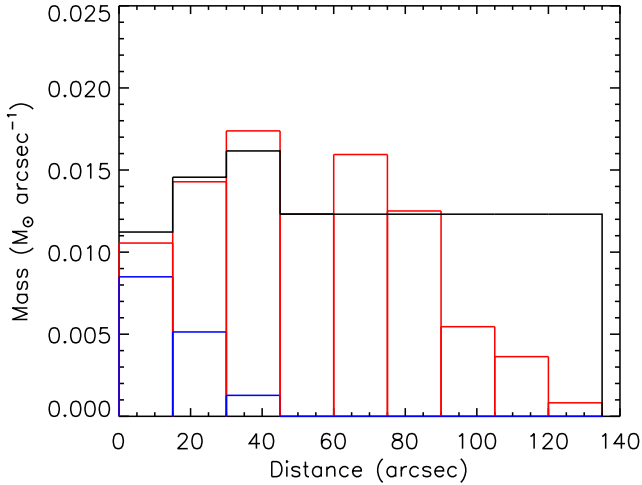}\\
\caption{Mass distribution of the outflow with respect to the distance from the central
source. The properties are integrated over several annuli shown in Figure \ref{fig:massmap}. The red and blue
curves represent the red lobe and the blue lobe. The black lines show the estimated mass distribution 
of the ambient material which originally filled the current outflow cavity.}
\label{fig:massdis}
\end{center}
\end{figure}

First, we study the question whether the molecular outflow is entrained locally
or entrained at small radii and then carried out.
If most of the outflow material is entrained locally, we expect that the mass distribution
of the outflow with respect to the distance from the central source is similar to the original mass distribution
of the ambient material which filled the outflow cavity.
The linear mass distribution of the outflow is calculated by 
integrating the mass surface density map of the outflow over semi-annuli centered at the central source
(the annuli are shown in the top panel of Figure \ref{fig:massmap}),
which are shown in Figure \ref{fig:massdis} with red and blue lines for the red and blue lobes. 
To estimate the mass distribution of the ambient gas which was filling the outflow cavity,
we assume the density profile of the original core is same as that of the current remaining core.
By fitting the sub-mm continuum emission, 
\citet[]{vanKempen09} found that the remaining core has $5~M_\odot$ within 0.1 pc ($46\arcsec$, 20800 AU) 
with the radial density profile following a power-law $\rho\propto r^{-1.8}$. 
Therefore, we assume the original material distribution in the outflow cavity 
to be composed of a cone with a half opening angle of $40^\circ$ with same density profile 
within $45\arcsec$ and 
a cylinder with a constant diameter and
a constant density of $10^4~\mathrm{cm}^{-3}$ from $45\arcsec$ to $120\arcsec$.
The shape of this simplified outflow cavity is shown in the first panel of Figure \ref{fig:massmap}.
The density of $10^4~\mathrm{cm}^{-3}$ for the outer part of the cloud is valid because
$\ceio$ and CS emission is detected throughout the region and it is consistent with the
the density profile for the inner core which gives $1.6\times 10^4~\mathrm{cm}^{-3}$ at $45\arcsec$.
\citet[]{vanKempen09} also estimated the density of the outer cloud to be a few $\times 10^3~\mathrm{cm}^{-3}$
from detection of $\twco$ 6-5 emission.
The estimated mass distribution of the ambient material which was filling the outflow cavity
is shown with black lines in Figure \ref{fig:massdis}. 
The figure shows that the outflow mass distribution is, in general, at an approximately similar level 
to the mass distribution
of the original core gas filling the outflow cavity, except in the regions close to the cloud edges
(outermost two annuli of the blue and red lobes) where the outflow mass drops fast with the distance.
This suggests that most of the outflow material is entrained locally from the ambient core, 
or not far from their current position, rather than being entrained close to the central source and then 
carried out. 
We note that, in the wide-angle wind entrainment model (see Section \ref{sec:outflow}), 
the outflowing shell contains
material accumulated all the way from the central source.
However, in such a case, the ``local entrainment'' is still valid in the sense that 
the ambient material joins the outflow at its distance from the source,
as opposed to infalling into the inner region first and then being launched and carried out.
The wide-angle wind entrainment also mainly works at the base of the outflow, 
as shown in Section \ref{sec:outflow}.

The above analysis implies that the core material be- comes part of the outflow as the outflow cavity broadens.
We then estimate the core destruction time scale based on this scenario of outflow broadening.
Assuming that the core material is being entrained into the outflow at a constant rate, 
i.e, the solid angle extended by the outflow cavity increases at a constant rate, 
the core destruction time scale can be estimated to be $\tau_\mathrm{des}=t_\mathrm{out}/(1-\cos \theta_c)$,
where $t_\mathrm{out}$ is the current age of the outflow, 
and $\theta_c$ is the half opening angle of the outflow cavity.
For the central 0.1 pc core, the current outflow opening angle is approximately $40^\circ$, 
then the total core destruction time is $\tau_\mathrm{des}=4 t_\mathrm{out}$.
There are several ways to estimate the age of the outflow.
The dynamical age of the {\it molecular outflow} is not a reliable estimate of its true age since
the outflow is entrained locally as discussed above.
The dynamical age of the parsec-scale jet (\citealt[]{Stanke99}) is about 9000 yr,
which is a lower limit since the jet could extend further.
Instead, we use the typical age of an early Class I source as the age of this outflow,
which is 0.13 to 0.26 Myr. This is estimated from the fact that the typical lifetime of 
the Class 0+I phase in low-mass star formation is 0.40 to 0.78 Myr, 
with about 1/3 of the time spent in the Class 0 phase (\citealt[]{Dunham14a}; \citealt[]{Dunham15}),
The core destruction timescale is then approximately 0.52 to 1 Myr, 
which is consistent with the aforementioned Class 0+I lifetime.
This time scale is also shorter than the collapsing time scale of the core (core mass divided by infall rate) 
which we estimate to be 1.5 Myr from the core mass of 5 $M_\odot$ and the current infall rate of 
$3.2\times 10^{-6}~M_\odot~\mathrm{yr}^{-1}$ estimated in Section \ref{sec:rotationcore}.
These suggest that the mass entrainment rate from the core to the outflow based on outflow broadening
is high enough for the outflow to potentially disperse the core within the time scale of the embedded 
phase of low-mass star formation.

The formation efficiency from core to star is a key parameter for setting the final mass of the star,
and it is believed to be strongly regulated by outflow feedback.
In a simple scenario, as the protostar grows, 
the material in the core either accretes onto the star (with a small fraction onto the disk) 
or is entrained by the outflow. 
Some material can be expelled from the core
by photoionizing winds, but this only becomes important for massive protostars.
Therefore we can define three efficiencies: 1) the instantaneous efficiency 
\begin{equation}
\epsilon(t)\equiv\frac{\dot{m}_*(t)}{\dot{m}_*(t)+\dot{m}_o(t)}, 
\end{equation}
where $\dot{m}_*$ is the accretion rate on to the protostar
and $\dot{m}_o$ is the rate at which the core material joins the outflow;
2) the current averaged efficiency,  
\begin{equation}
\bar{\epsilon}(t)\equiv\frac{\int_0^t\dot{m}_* dt}{\int_0^t(\dot{m}_*+\dot{m}_o) dt}=
\frac{m_*(t)}{m_*(t)+m_o(t)},
\end{equation} 
where $m_*(t)$ is the current protostellar mass and $m_o(t)$ is the current outflow mass;
and 3) the final efficiency, 
\begin{equation}
\bar{\epsilon}_f\equiv \bar{\epsilon}(t_f)=
\frac{m_*(t_f)}{m_*(t_f)+m_o(t_f)}=\frac{m_{*,f}}{M_c},
\end{equation} 
where $t_f$ is the formation time, $m_{*f}$ is the final mass of the protostar, and $M_c$ is the initial mass of the core.
For the instantaneous core-to-star efficiency, 
the protostellar accretion rate can be approximated by the infall rate
of the innermost envelope, which is $3.2\times 10^{-6}~M_\odot~\mathrm{yr}^{-1}$
(Section \ref{sec:rotationcore}).
On a similar scale (within $6\arcsec$), 
the outflow material with a velocity higher than the 
escaping velocity has a total momentum of 0.07 $M_\odot~\kms$, which 
corresponds to an instantaneous mass loading rate of 
$0.07~M_\odot~\kms~/6\arcsec=5.5\times 10^{-6}~M_\odot~\mathrm{yr}^{-1}$ for the outflow.
The ratio between the infall rate and the outflow mass loading rate indicates the  
instantaneous core-to-star efficiency is about 1/3.
For the current average core-to-star efficiency,
by fitting the position-velocity diagrams of the $\thco$ and $\ceio$ flattened structure surrounding the
central source, we find the dynamical mass of the central source is $0.3~M_\odot$
 (see Section \ref{sec:rotationcore}).
Within $45\arcsec$ (i.e. the size of the 5 $M_\odot$ core), 
the total outflow mass is 0.85 $M_\odot$, and the total mass of the outflowing material with a velocity higher
than the escape velocity is 0.6 $M_\odot$,
which are 2 to 3 times the mass of the protostars.
This corresponds to a current average core-to-star efficiency 
of 1/3 to 1/4.
These estimates on current core-to-star efficiencies are consistent with the final efficiencies estimated from 
observations of CMF and IMF or results of theoretical simulations (e.g. \citealt[]{Federrath14}, \citealt[]{Offner14a}).
Thus, it appears that the outflow is already significantly influencing the star formation efficiency 
towards the driving protostar of the HH46/47 outflow.

\begin{figure*} 
\begin{center}
\includegraphics[width=0.9\textwidth]{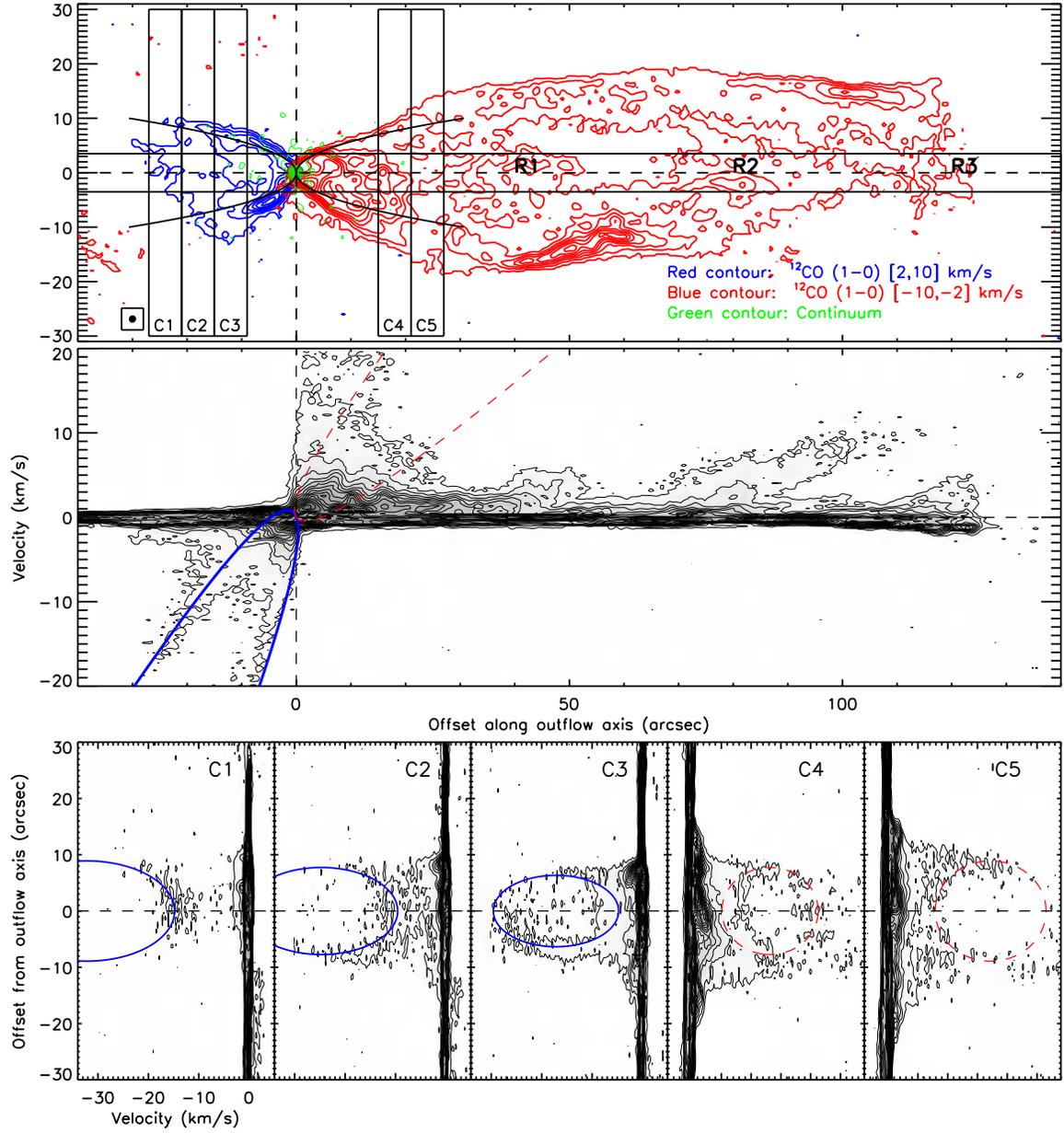}\\
\caption{{\bf Top:} $\twco$ (1-0) emission integrated from 2 to $10~\kms$ 
shown in the red contours, and emission integrated from $-10$ to $-2~\kms$ 
shown in the blue contours. The images are rotated by 30$^\circ$ counterclockwise. 
The contours start at $3\sigma$ with a step of $6\sigma$ ($1\sigma=21~\mJybeam~\kms$).
The rectangles show the cuts for the PV diagrams shown in the panels below. 
The parabolas show the projected shape of the model outflow cavity.
{\bf Middle:} The position-velocity diagram of $\twco$ (1-0) along the outflow axis with a cut width of 7$\arcsec$. 
The contours start at 38 $\mJybeam$ with a step of 38 $\mJybeam$.
The blue and red curves show the PV diagrams from a model where outflow shells are driven by a wide-angle wind
(see Section \ref{sec:outflow}).
{\bf Bottom:} Position-Velocity diagrams of $\twco$ (1-0) along 6$\arcsec$-wide cuts perpendicular to the outflow
axis. The contours start at 36 $\mJybeam$ with a step of 36 $\mJybeam$. The blue and red ellipses show
the model fit.}
\label{fig:pv_12CO}
\end{center}
\end{figure*}

\begin{figure*} 
\begin{center}
\includegraphics[width=\textwidth]{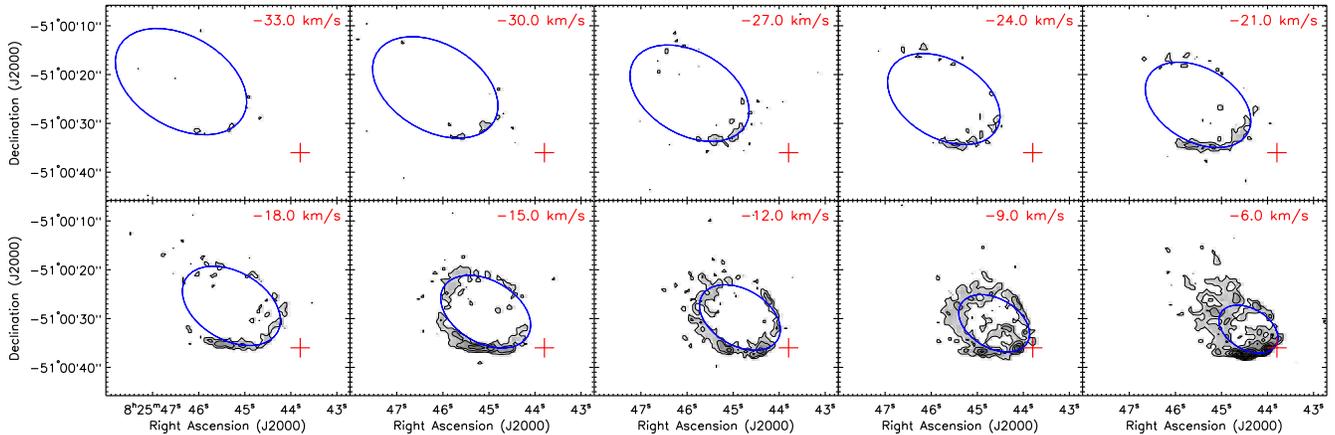}\\
\caption{Channel maps of the $\twco$ (1-0) emission at velocities less than or equal to 
$-6~\kms$ showing the blue-shifted
outflow. The channel width is 3 $\kms$. The contours start at 3$\sigma$ with
a step of 6$\sigma$ ($1\sigma=4.5~\mJybeam$).
The synthesized beam is $1.35\arcsec \times 1.30\arcsec$ (P.A. = $-55.2^\circ$).
The red crosses mark the central source (peak of the continuum emission). The blue ellipse in each panel shows
the expected shape of emission from the model where outflow shells are driven by a wide-angle wind 
(see Section \ref{sec:outflow}).}
\label{fig:chan_blue}
\end{center}
\end{figure*}

\begin{figure*} 
\begin{center}
\includegraphics[width=0.7\textwidth]{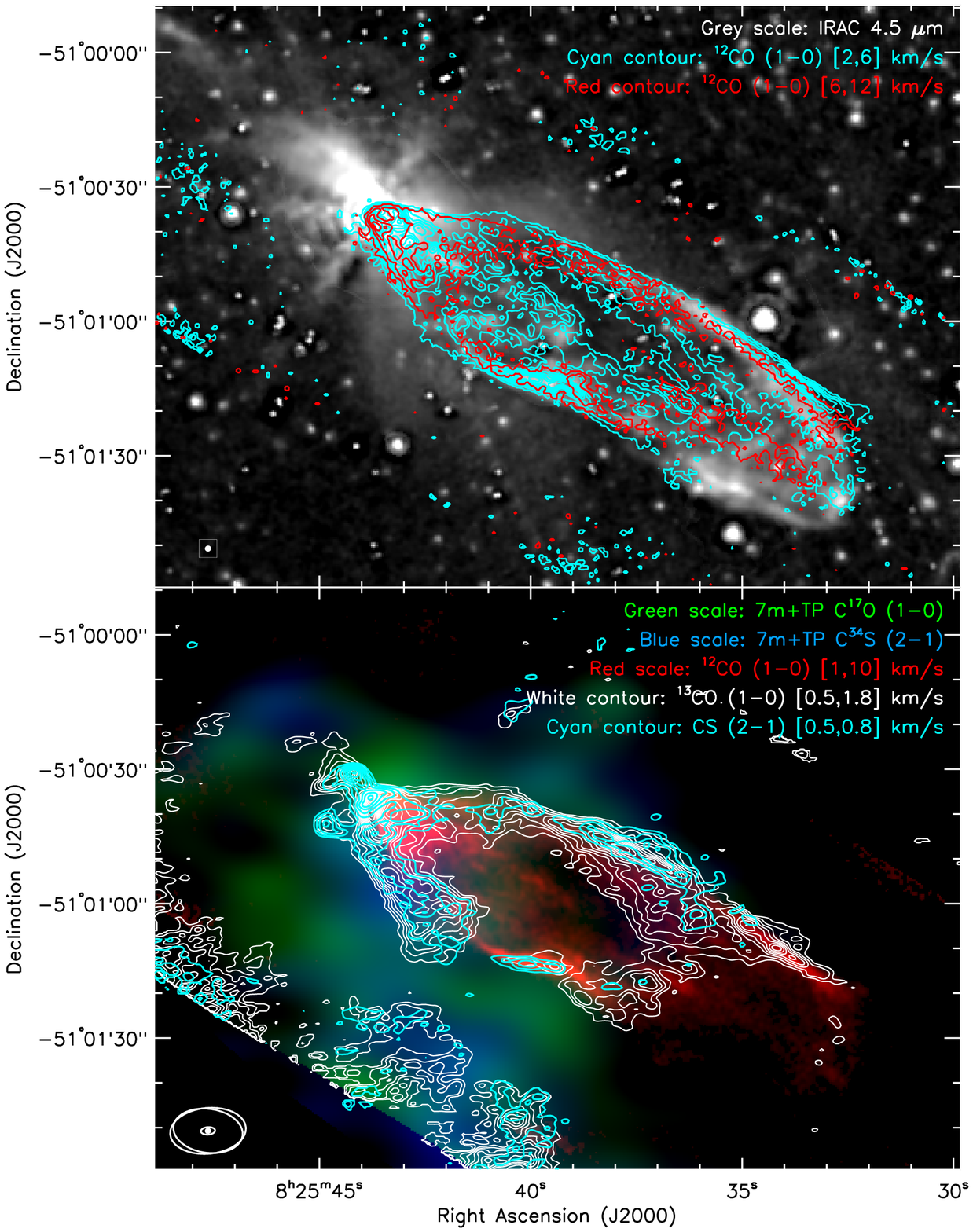}
\caption{{\bf Upper panel:} Comparison of the red-shifted CO outflow with the {\it Spitzer} IRAC 4.5 $\micron$ image.
The {\it Spitzer} data are from \citet[]{Noriega04} and have been reprocessed
with a deconvolution algorithm to reach an angular resolution of about $0.6\arcsec$ to $0.8\arcsec$,
with 60 iterations (see \citealt[]{Noriega12} for details). 
The cyan and red contours show the $\twco$ emission integrated from 1 to $6~\kms$
and from 6 to $12~\kms$ respectively.
The lowest contour and subsequent contour steps are $3\sigma$ and $6\sigma$ 
($1\sigma=15~\mJybeam~\kms$ for the cyan contours and $1\sigma=19\mJybeam~\kms$ for the red contours).
{\bf Lower panel:} The red-shifted outflow cavity traced by $\twco$, $\thco$ and CS.
The red color scale shows the integrated emission of $\twco$ (1-0) from 1 to $10~\kms$. The
white and cyan contours show the $\thco$ (1-0) emission integrated from 0.5 to $1.8~\kms$ and 
the CS (2-1) emission integrated from 0.5 to $0.8~\kms$.
The lowest contour and subsequent contour steps are $6\sigma$ and $3\sigma$, respectively
($1\sigma=6.4~\mJybeam~\kms$ for white contours and $1\sigma=2.2~\mJybeam~\kms$ for cyan contours).
The integrated emission of $\cseo$ (1-0) and $\cths$ (2-1) are also shown for reference 
in green and blue color scales, respectively.}
\label{fig:int_IRAC}
\end{center}
\end{figure*}

\begin{figure*} 
\begin{center}
\includegraphics[width=0.9\textwidth]{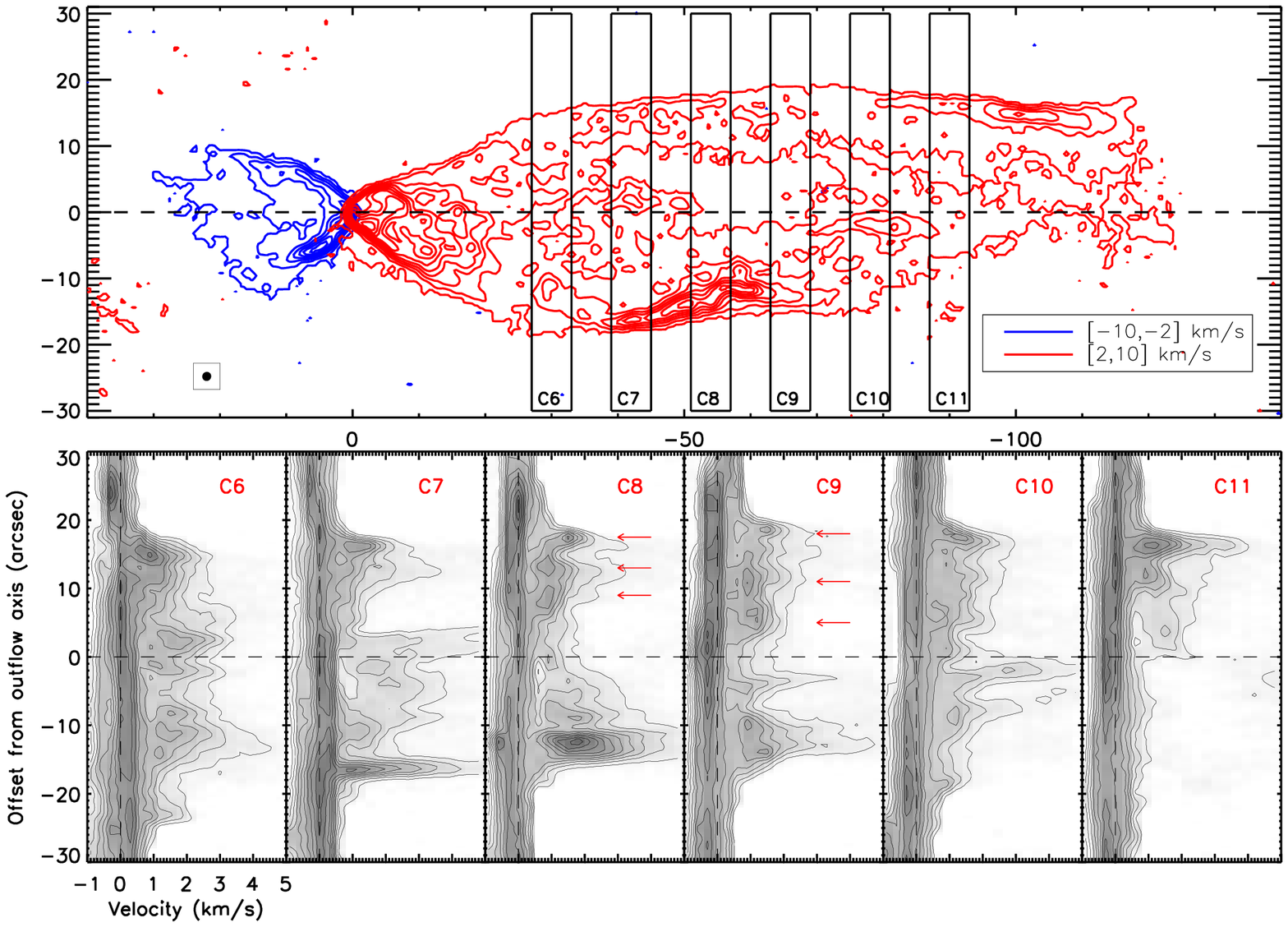}\\
\caption{{\bf Upper panel:} 
Integrated intensity maps of the $\twco$ red and blue-shifted outflow lobes 
(the same as the top panel of Figure \ref{fig:pv_12CO}).
{\bf Lower panel:} The position-velocity diagrams of $\twco$ (1-0) along 6$\arcsec$-wide cuts perpendicular to the outflow
axis. The contours start at 36 $\mJybeam$ with a step of 36 $\mJybeam$.}
\label{fig:pv_12CO_red}
\end{center}
\end{figure*}

\subsection{Mechanisms of Outflow Entrainment }
\label{sec:outflow}

In Paper I, it was argued that the blue-shifted outflow is mainly entrained by a wide-angle wind and
the red-shifted outflow is mainly entrained by the jet. Here we revisit the question of entrainment mechanism
with the new ALMA data which has better angular resolution, 
recovers more of the extended emission and includes more tracers.

\paragraph{Blue-shifted outflow}

The kinematics and morphology of the blue-shifted outflow at velocities higher than about $6~\kms$
can be explained by a model in which the molecular outflow is swept up by a 
wide-angle wind.
For a radial wind with force $\propto 1/\sin^2(\theta)$ interacting with a flattened ambient core with 
density $\propto \sin^2(\theta)/r^2$ and instantaneously mixing with shocked ambient gas, 
the swept-up shell is a radially expanding parabola with a 
Hubble law velocity structure (\citealt[]{Li96}; \citealt[]{Lee00}).
Following the simple analytical description by \citet[]{Lee00}, the morphology of such an outflowing shell 
is described by a parabola in the form of $z=CR^2$, with the $z$-axis along the outflow axis and the 
$R$-axis perpendicular to it,
and the velocities of the shell on the directions of $z$ and $R$ are described by $v_z=v_0 z$ and $v_R=v_0 R$.
The free parameters in such a model are the inclination $i$ between the outflow axis
and the plain of the sky, $C$, and $v_0$.

As shown in Figures \ref{fig:pv_12CO} and \ref{fig:chan_blue}, such a model successfully reproduces
the features in the $\twco$ position-velocity (PV) diagrams along and perpendicular 
to the outflow axis and the $\twco$ channel maps of the blue-shifted outflow 
at velocities higher than about $6~\kms$. 
Along the outflow axis, the PV diagram of the blue-shifted outflow shows an inclined parabolic structure.
Meanwhile in the PV diagrams perpendicular to the outflow axis, elliptical structures are seen.
The channel maps at velocities $\le -6~\kms$ also show elliptical ring structures. 
All these features are well fitted with our simple analytical model. 
The parameters of the best-fit model is $i=40^\circ\pm 1^\circ$, $C=0.4\pm0.1~\mathrm{arcsec}^{-1}$, 
and $v_0=1.5\pm0.1~\kms~\mathrm{arcsec}^{-1}$. 
The uncertainties are the range of the values with which
relatively good fits can be achieved by visual inspection.
The fitted inclination of the outflow is consistent with the value derived from the observations
of the optical jet by \citet[]{Eisloffel94} and \citet[]{Hartigan05}, which are $34^\circ\pm3^\circ$ and
$37^\circ.5\pm2^\circ.5$ respectively.
The parameter $v_0$ corresponds to a time scale of
$t_0=1/v_0=1.4\times 10^3$ yr, which can be considered as the dynamical age of the 
wide-angle-wind entrained outflow. 
This age is even shorter than the dynamical age of the jet (9000 yr) and is certainly
much shorter than the expected age for an outflow driven by an early Class I source 
(several $\times 10^5$~yr, see discussion in Section \ref{sec:compare}). This suggests that
the wide-angle wind entrainment has started or began to be visible only very recently.

The very young age of the wide-angle wind entrained outflow is consistent with the fact
that the majority of the blue-shifted outflow emission is at lower velocities and not following
the prediction of the wide-angle wind model.
At a lower velocity, the emission is concentrated along the parabolic outflow cavity walls (see
Figure \ref{fig:chan_12CO}), which may be the material that was entrained before and has slowed down 
while interacting with the remaining core. Jet bow-shock entrainment, even though there is no clear evidence
left in the morphology or kinematics of the emission, could still be responsible for entraining these material in the past,
considering an optical jet is seen inside the blue-shifted outflow cavity.
The fitted power-law index of the mass-velocity relationship of the blue-shifted outflow 
(Figure \ref{fig:massspec_out1}, Section \ref{sec:mass})
is more consistent with the jet bow-shock entrainment model, which predicts $-3.5<\gamma<-1.5$, than the wide-angle
wind entrainment model, which predicts $-1.8<\gamma<-1.3$ (e.g. \citealt[]{Lee01}). 
In fact, fitting a power-law for the high-velocity portion ($|\vout|>6~\kms$) gives an index of $\gamma=-1.27$,
consistent with the wide-angle wind model, and fitting a power-law for the low-velocity portion ($|\vout|<6~\kms$)
gives an index of $\gamma=-3.74$ which is more consistent with the jet bow-shock model.
This again suggests that two entrainment mechanisms coexist in the blue-shifted outflow.

\paragraph{Red-shifted outflow}

The kinematics and morphology of the red-shifted outflow show evidences of jet bow-shock entrainment.
Figure \ref{fig:pv_12CO} shows that the three $\twco$ clumps R1, R2 and R3 on the red-shifted outflow axis
have distinctive kinematic features that the velocity increases with the distance to the central source.
This type of feature is generally called ``Hubble wedges'', 
and is considered to be produced by the entrainment of ambient gas by the jet bow-shock (\citealt[]{Arce01a}).
The three $\twco$ clumps are also coincident with the shocked region shown as bright IR or optical knots 
along the red-shifted outflow axis (Figure \ref{fig:int_IRAC}, upper panel), 
or at same distances with respect to the central source 
as the bright knots in the blue-shifted optical jet, which also trace shocks.
Therefore the $\twco$ clumps R1, R2 and R3 mark the positions of the apices of three bow-shocks 
produced by episodic mass ejection, as argued in Paper I.

The three $\twco$ clumps are connected with the outflow cavity walls by diffuse emission,
following bow-shock shapes (Figure \ref{fig:pv_12CO}, top panel).
More careful inspection reveals that the northern outflow cavity wall is actually composed of
multiple shells associated with these bow-shocks.
As mentioned in Section \ref{sec:12CO}, two or more parallel thin structures can be identified
in the northern wall of the outflow cavity from about 30$\arcsec$ from the central source 
to the end of the red lobe in the channel maps at velocities from $1.8$ to $3.6~\kms$.
The inner layer of the cavity wall deviates from 
the outer layer and curves towards clump R2 on the axis
at a distance of about $90\arcsec$ from the central source, 
while the outer layer extends further out.

We can also identify these different layers in the outflow cavity wall in the PV diagrams.
Figure \ref{fig:pv_12CO_red} shows the PV diagram perpendicular to the outflow axis at several 
positions along the red lobe. The PV diagram in the C9 cut shows that the northern outflow
cavity wall is composed of three components at distances of approximately $5\arcsec$, $10\arcsec$
and $18\arcsec$ from the axis (marked with red arrows). 
Each of these components has an emission peak and wider velocity range
than the emission between them. Such structures are seen consistently over the red-shifted outflow
as shown in Figure \ref{fig:pv_12CO_red}. 
At a further distance from the source (cut C10), the innermost components
of the northern outflow cavity wall seen in the cut C9 move towards the axis and start to merge with
the emission of clump R2. At a distance closer to the central source (cut C8), this inner component
is further away from the axis.

The different shells are also seen in different velocities and tracers.
In the upper panel of Figure \ref{fig:int_IRAC}, close to the end of the outflow cavity, 
the high velocity emission follows the infrared emission and curves
to the outflow axis, tracing the bow-shock whose apex is at the end of the IR outflow cavity
(i.e. the $\twco$ clump R3).
On the other hand, part of the low velocity emission remains parallel to the outflow axis and 
deviates from the IR emission, suggesting it may be material entrained by bow shocks which have 
already moved out of the cloud, which is consistent with the fact that the jet extends to a parsec scale distance.
In the lower panel of Figure \ref{fig:int_IRAC}, most of the low velocity $\thco$ and CS emission trace a closed outflow cavity
structure which ends around the position of the R2 clump,
while some of the $\thco$ emission extends further away to the position of 
clump R3.

All these features suggest that the extended red-shifted outflow (with distance $\gtrsim 30\arcsec$ from
the central source) is composed of several nested shells formed by entrainment of a series of jet bow-shocks.
Such a scenario is consistent with theoretical models of jet bow-shock entrainment (e.g.
\citealt[]{Raga93}; \citealt[]{Lee01}) and is also supported by recent observations in other sources (e.g.
HH212, \citealt[]{Lee15}). Here with the unprecedented angular resolution and sensitivity of ALMA, 
we not only identify the shells around the apices of the bow-shocks but also resolve them in the outflow cavity
walls.

However, at the base of the outflow cavity, the wide-angle wind may still contribute.
For example, Figure \ref{fig:pv_12CO} shows that some of the PV diagrams perpendicular to the outflow axis
also show elliptical rings (Lower panel, cut C4 and C5) similar to those seen on the blue lobe at the same 
distances to the central source. The red dashed ellipses in the figure show the fits to these features 
using the wide-angle wind entrainment model described above,
with the same inclination $i$ and outflow cavity shape $C$, but a slower velocity, $v_0=0.8~\kms~\mathrm{arcsec}^{-1}$. 
However, we do not find clear evidence of a wide-angle-wind-entrained outflow 
in the PV diagram along the outflow axis, although some of the high
velocity emission close to the central source seems to be consistent with such a model. 
Also the $\twco$ channel maps do not show the ring structures expected by the wide-angle wind
model as in the blue-shifted outflow.
We also see a wider structure at the base of the red-shifted outflow 
which can be evidence of a wide-angle wind. 
It is most clearly seen in the $\twco$ channel maps at velocities from 0.6 to $2.1~\kms$, 
especially towards the south of the central source. A similar, but even wider structure appears
in low velocity $\thco$ and CS emission (Figure \ref{fig:int_IRAC}, lower panel). 
Spatially these coincide
with the diffuse emission in the IRAC 4.5 $\micron$ continuum images (Figure \ref{fig:int_IRAC}, upper panel).
These extended IR emissions were argued to be scattered light by an outflow cavity
wider than that in IR shock emissions (\citealt[]{Velusamy07}). Therefore the structure
in low velocity CO and CS emission may be tracing the outflowing material in this wider cavity,
which is entrained by a wide-angle wind. At a larger polar angle from the outflow axis,
the wide-angle wind is slower and therefore the 
entrained material only appears in the low velocity range.
In addition, as discussed in Section \ref{sec:mass}, the energy map of the outflow 
(Figure \ref{fig:massmap}, lower panel)
shows that, in the red lobe, the kinetic energy is concentrated at the base of the outflow cavity, which is not
consistent with a jet bow-shock entrainment scenario in which 
most of the energy is expected to be injected at the heads of bow-shocks,
suggesting a second entrainment mechanism (such as wide-angle wind) is in action.
~\\

To summarize, we find evidence that jet bow-shock entrainment and wide-angle wind entrainment co-exist 
in both the blue-shifted and red-shifted outflows, 
although which mechanism is more visible differs on the two sides.
While the blue-shifted outflow cavity has been cleared and little material is along the jet path,
making the wide-angle wind entrainment apparent, the red-shifted outflow cavity has a large amount of remaining 
dense material resulting in significant jet bow-shock entrainment in this region.
Since the outflow cavity is gradually cleared as a protostar evolves, we would expect to see the jet
entrainment is more visible in an earlier stage while the wide-angle wind {becomes visible} in a later stage.
We note that the jet and wide-angle wind do not need to be two distinct wind types, rather they
can be two components of a single wind from the accretion disk with its density and/or velocity depending on 
the polar angle from the outflow axis (e.g. \citealt[]{Cabrit99}; \citealt[]{Shang06}).

\subsection{Rotational Structure Around the Central Source}
\label{sec:rotationcore}

\begin{figure*} 
\begin{center}
\includegraphics[width=0.7\textwidth]{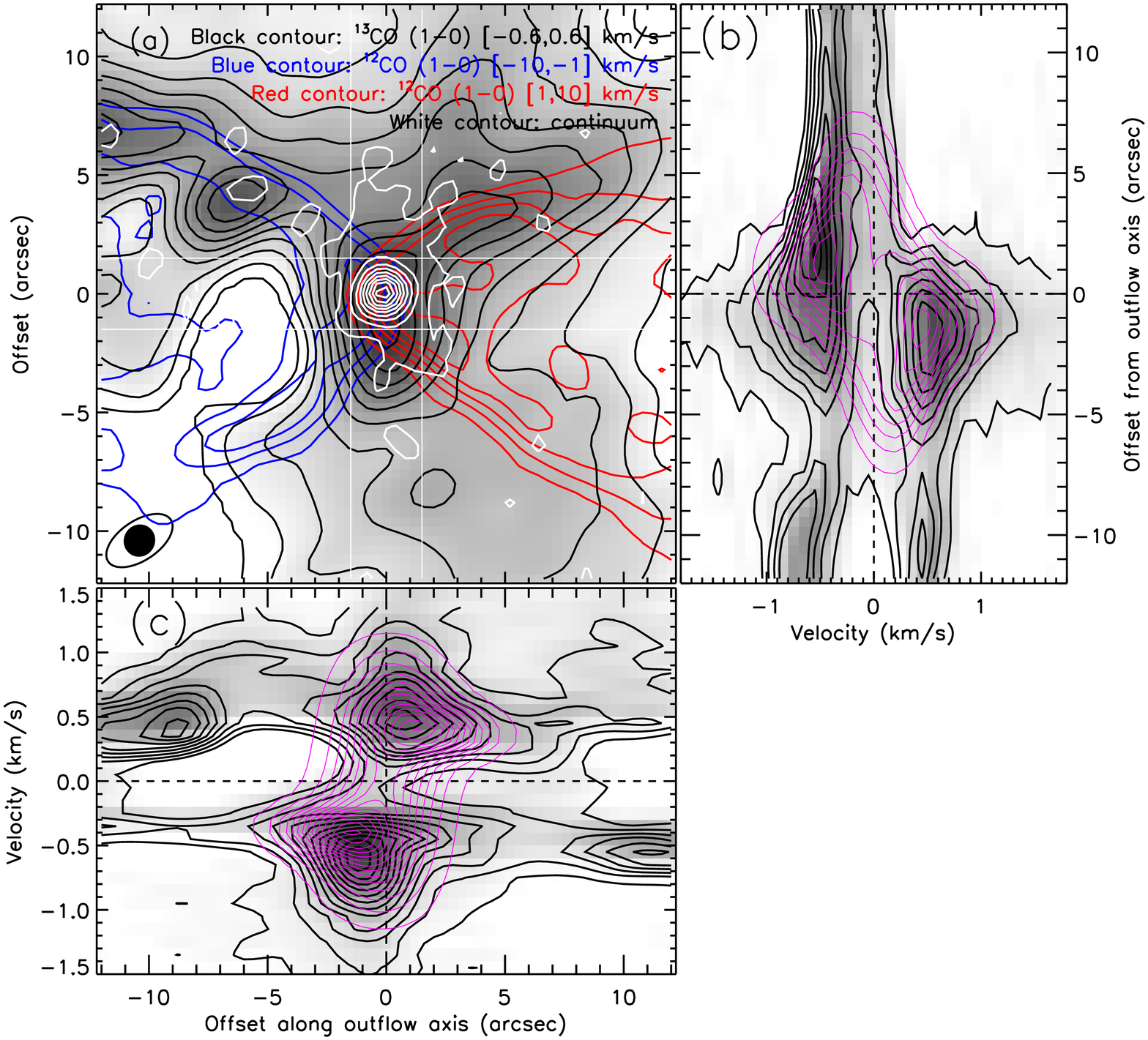}\\
\caption{(a) $\thco$ emission integrated from $-0.6$ to $0.6~\kms$ (grey scale and black contours) 
showing the envelope. Only the interferometric data is used.
The contours start from $3\sigma$ with a step of $6\sigma$ 
($1\sigma=6.2~\mJybeam~\kms$). The red and blue contours show the red-shifted (1 to $10~\kms$) 
and blue-shifted ($-10$ to $-1~\kms$)  $\twco$ outflows for reference. 
The white contours, which show the continuum emission, start at $3\sigma$ and have a step of
$15\sigma$ ($1\sigma=0.04~\mJybeam$). The images are rotated counterclockwise
by $30^\circ$ so that the outflow axis lies along the x-axis.
The white stripes show the cuts used for the PV diagrams.
(b) The position-velocity diagram of $\thco$ (1-0) along the 3$\arcsec$-wide cut 
perpendicular to the outflow axis (black contours and grey scale). The purple contours
show the PV diagram expected from a model including infall and rotation.
Both black and purple contours start at 36 $\mJybeam$ with a step of 72 $m\Jybeam$. 
(c) Same as panel (b) but along the outflow axis. 
The contours start at 24 $\mJybeam$ with a step of 48 $\mJybeam$.}
\label{fig:rot_13CO}
\end{center}
\end{figure*}

\begin{figure*} 
\begin{center}
\includegraphics[width=0.7\textwidth]{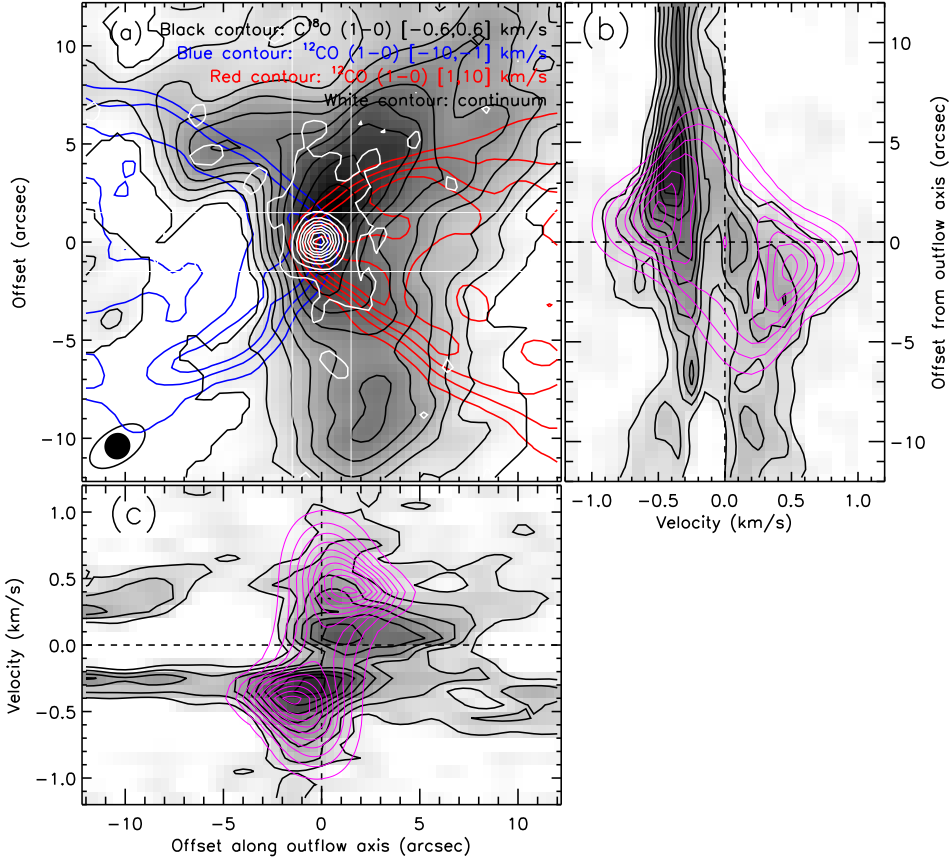}\\
\caption{(a) Same as Figure \ref{fig:rot_13CO}(a), but
with the grey scale and black contours showing the $\ceio$ emission integrated from $-0.6$ to $0.6~\kms$.
The contours start from $3\sigma$ with a step of $3\sigma$ ($1\sigma=5.6~\mJybeam~\kms$). 
(b) Same as Figure \ref{fig:rot_13CO}(b), but for $\ceio$.
Both black and purple contours start at 33 $\mJybeam$ with a step of 33 $\mJybeam$. 
(c) Same as Figure \ref{fig:rot_13CO}(c), but for $\ceio$.
The contours start at 27 $\mJybeam$ with a step of 27 $\mJybeam$.}
\label{fig:rot_C18O}
\end{center}
\end{figure*}

\begin{figure*} 
\begin{center}
\includegraphics[width=0.9\textwidth]{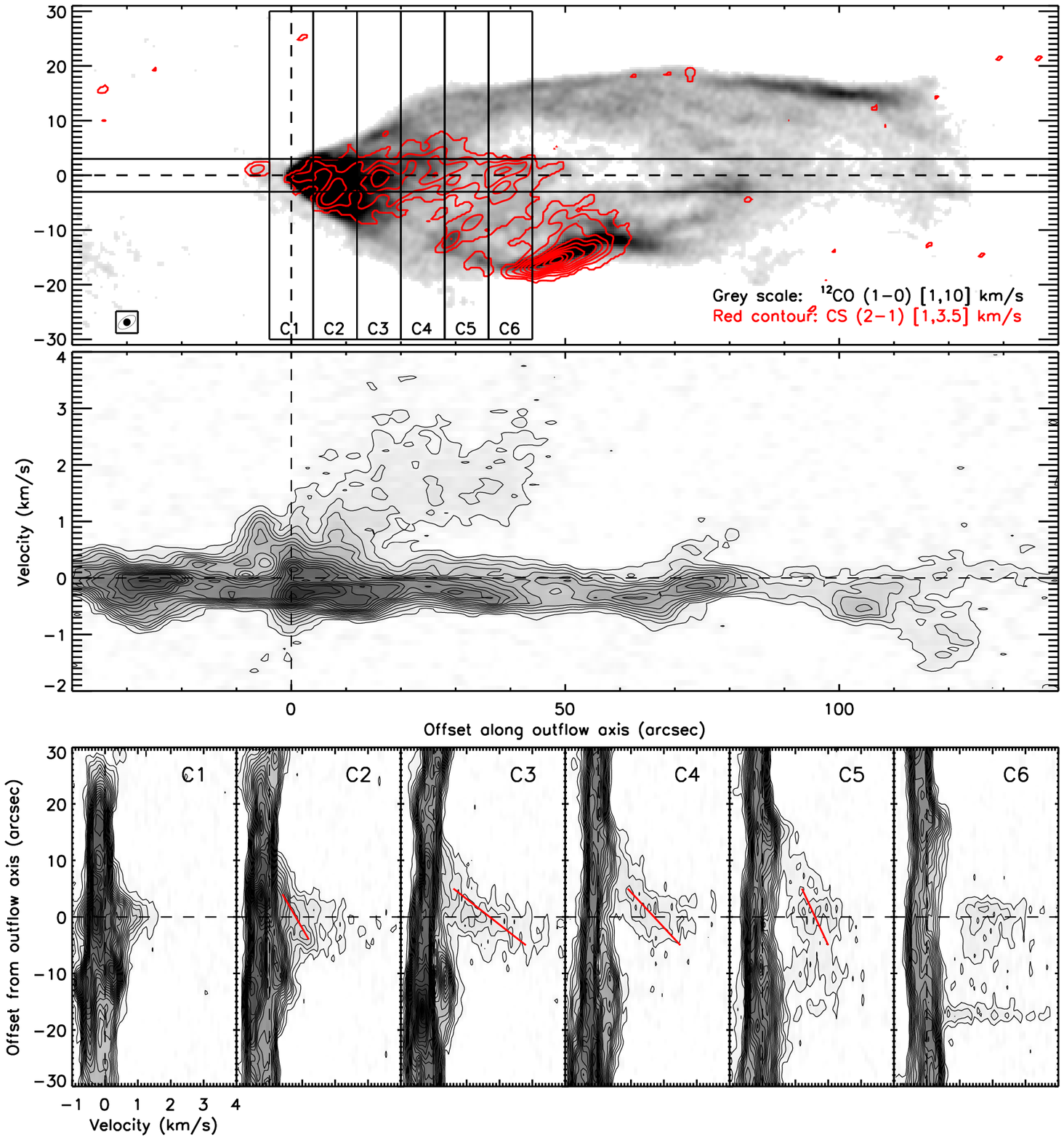}\\
\caption{{\bf Top:}  CS (2-1) emission integrated from 1 to $3.5~\kms$ relative to
the cloud velocity shown in red contours
overlaid on $\twco$ (1-0) image integrated from 1 to $10~\kms$ in grey scale. 
The images are rotated by 30$^\circ$ counterclockwise. The contours start at $3\sigma$ with
a step of $3\sigma$ with $1\sigma=6.5~\mJybeam~\kms$. The cuts for the PV diagrams are shown. 
{\bf Middle:} Position-Velocity diagram of CS (2-1) along the outflow axis with a cut width of 6$\arcsec$. 
The contours start at 48 $\mJybeam$ with a step of 48 $\mJybeam$.
{\bf Bottom:} Position-Velocity diagrams of CS (2-1) along 8$\arcsec$-wide cuts perpendicular to the outflow
axis. The contours start at 36 $\mJybeam$ with a step of 36 $\mJybeam$. The red lines indicate the velocity
gradients across the outflow axis.}
\label{fig:pv_CS}
\end{center}
\end{figure*}

As mentioned in Sections \ref{sec:13CO} and \ref{sec:C18O},
velocity gradients are found across the central source perpendicular to the outflow axis
in both $\thco$ and $\ceio$, indicating a rotating structure around the central source.
In Figures \ref{fig:rot_13CO} and \ref{fig:rot_C18O}, the $\thco$ and $\ceio$ integrated emissions
show a flattened structure around the central source 
with its major axis perpendicular to the outflow axis.
Its size is about $10\arcsec$ (4500 AU) across, which is much larger than 
what is expected for a rotationally supported Keplerian disk 
(typically few hundred AU for Class 0 and I sources; e.g. \citealt[]{Yen13}). 
Therefore it is likely to be a rotating envelope that feeds the accretion disk which we are not able to resolve
with our current data.
The morphology of this flattened structure is affected by the outflow.
On the blue-shifted side, its boundary outlines the outflow cavity. 
On the red-shifted side, the brightest part of the flattened structure bends following
the red-shifted outflow cavity.
The peaks of the $\thco$ and $\ceio$ integrated emissions are close to the continuum peak. 
The most extended continuum emission is elongated in the north-south direction, 
consistent with the brightest parts of the $\thco$ and $\ceio$ emissions.

The upper-right panel of Figure \ref{fig:rot_13CO} shows the PV diagram of $\thco$ 
along the major axis of the flattened structure (i.e. perpendicular to the outflow axis).
In addition to the two major components within a distance of about $6\arcsec$ from the central source,
there is emission at $-1$ to $-0.5~\kms$ and at $0.5~\kms$ at
a position between $-8$ to $-12\arcsec$ from the center. The blue-shifted emission across this
region is associated with the larger cloud structure at this velocity (see Figure \ref{fig:chan_13CO}),
and the red-shifted component is separated from the central flattened structure 
and appears to come from the outflow.
Therefore we only focus on the two main components within 6$\arcsec$ of the center on the PV diagram.
At higher velocities, the emission is confined near the central source (i.e. within 2$\arcsec$), 
while at lower velocities 
the emission extends to the north and south of the source. Although the blue-shifted and red-shifted 
emission peaks lie on different sides of the central source, the fainter emission is more symmetric.
The PV diagram clearly shows a signature of rotation around the central source, but the high velocity emission
across the rotation axis suggests there is also infall or outflow motion involved.
The lower-left panel of the same figure shows the PV diagram of $\thco$ along the minor axis
of this flattened structure (i.e. along the outflow axis). We also can identify two main components showing
a velocity gradient along the direction of the outflow axis, which again indicates that infall or outflow motion
exists in addition to rotation. 
Figure \ref{fig:rot_C18O} shows similar features for the $\ceio$ flattened envelope. 

We compare the observed PV diagrams with a simple analytic model 
similar to those used by \citet[]{Ohashi97} and \citet[]{Lee06}. 
In this model, the rotation velocity is assumed to be inversely proportional to the radius, therefore
conserving angular momentum, $v_\mathrm{rot}=j/R$, and the infalling
velocity is then $v_\mathrm{infall}=\sqrt{2Gm_*/R-(j/R)^2}$ as imposed by mechanical energy conservation
in the potential well of the central source (e.g. \citealt[]{Sakai14}).
The envelope extends from the centrifugal barrier radius
where all kinetic energy is converted to the rotational motion
$R_\mathrm{in}=j^2/(2Gm_*)$ to an outer radius $R_\mathrm{out}$ with a constant thickness of $H$. 
The density and temperature distributions follow power-laws $n\propto R^{-k_\rho}$ and 
$T=T_{1000\mathrm{AU}}(R/1000~\mathrm{AU})^{-k_T}$, where $T_{1000\mathrm{AU}}$ is the temperature
at $R=1000$~AU. Such a model has 9 parameters $j$, $m_*$, $R_\mathrm{out}$, $H$, $k_\rho$, 
$m_\mathrm{env}$, $k_T$,  $T_{1000\mathrm{AU}}$ and inclination $i$, 
where the mass of the envelope $m_\mathrm{env}$, together with $R_\mathrm{out}$ and $H$,
is used to set the density. We try to use a single set of the parameters to reproduce both the $\thco$ and
$\ceio$ PV diagrams.
To narrow our search, we fix $R_\mathrm{out}=6\arcsec$ (2700 AU) from the current observation,
and $k_\rho=1.8$ from the observation by \citet[]{vanKempen09}. The temperature
is assumed to follow the dust temperature profile by heat from a 12 $L_\odot$ protostar (Eq. 2 in \citealt[]{Motte01}),
with $k_T=0.4$ and $T_\mathrm{1000AU}=25$~K.
The purple contours in Figures \ref{fig:rot_13CO} and \ref{fig:rot_C18O} show an example of our model. 
The fitted parameters are the specific angular momentum $j =1~\mathrm{arcsec}~\kms$ (450 AU $\kms$), 
the dynamical central mass $m_*=0.3~M_\odot$, the envelope mass $m_\mathrm{env}=0.1~M_\odot$,
the envelope thickness $H=1\arcsec$ (900 AU), and the inclination
between the line of sight and the envelope mid-plane $i=30^\circ$.
We are not performing a detailed model fit here, but rather we aim to show that the 
observed PV diagrams of the flattened structure are consistent 
with a collapsing envelope with rotation. 
The model reproduces most of the features in the PV diagrams, including the velocities, positions, and
intensities of the emission peaks, and the different behaviors of the emissions at high and low velocities.
However, the observation shows emission more spread out in space than the model. 
We also note that the emission peaks in the $\ceio$ PV diagram along the minor axis of the envelope 
are not symmetric in velocity, which may be caused by the outflow. 

Better agreement between the model and the observation can be achieved if we use different parameters
for $\thco$ and $\ceio$. The PV diagrams of $\thco$ can be better fitted with $j=1.2~\mathrm{arcsec}~\kms$
and $m_*=0.4 - 0.5~M_\odot$, and the PV diagrams of $\ceio$ can be better fitted with 
$j=1~\mathrm{arcsec}~\kms$ and $m_*=0.2~M_\odot$. The differences between these models can be considered
as the uncertainties of these parameters. 
Compared to the parameters describing the dynamics ($j$, $m_*$), the parameters that describe the geometry or
density distribution ($H$, $m_\mathrm{env}$) are less well constrained. 
The inclination of the model envelope, 
which is mainly constrained by the position separation of the emission peaks on the two PV diagrams,
differs from the outflow inclination (Section \ref{sec:outflow}).
These all suggest that the geometry or density/temperature distribution of the envelope in this
model is over-simplified.
Note that here we neglect the effect of CO depletion, which is expected 
at temperature $\lesssim 20$~K in protostellar envelopes (e.g. \citealt[]{Zhang15}), 
which happens at about $R=1800$~AU (4$\arcsec$) for the temperature profile we adopt.
Therefore it is likely that we have underestimated the envelope mass. 

From the estimated specific angular momentum $j$ and the central mass $m_*$,
the corresponding centrifugal barrier radius is $R_\mathrm{in}=0.85\arcsec$ (380 AU),
which can be considered as the outer radius of a rotationally supported Keplerian disk.
This size is at the high end of typical Class 0 and I disks (\citealt[]{Harsono14}).
The mass of the envelope $m_\mathrm{env}=0.1~M_\odot$ is consistent with the mass traced by
$\thco$ (0.09 $M_\odot$ with $\Tex=15$ K) and $\ceio$ (0.1 $M_\odot$ with  $\Tex=15$ K)
within a $12\arcsec \times 6\arcsec$ rectangle centered on the central source, using only the interferometric data.
This mass is likely a lower limit of the envelope mass since CO may be depleted in cold region.
On the other hand, the dust mass of 0.3 $M_\odot$ estimated from the continuum emission (Section \ref{sec:contin})
provides an upper limit of the envelope mass, since a large fraction of the continuum emission is from
the central peak which may trace the unresolved disk rather than the infalling envelope.
From $j$ and $m_*$ estimated above,
the infalling velocity is $0.41~\kms$ at a distance of $6\arcsec$ (2700 AU).
We then estimate the infall rate to be 
$\dot{m}_\mathrm{infall}=m_\mathrm{env}v_\mathrm{infall}/R=3.2 \times 
10^{-6}~M_\odot~\mathrm{yr}^{-1}$ with $m_\mathrm{env}=0.1~M_\odot$.
Note that the infall rate here is still an lower limit because of the neglect of CO depletion.
 \citet[]{Hartigan94} estimated the mass loading rate of the jet to be 
$4\times 10^{-7}~M_\odot~\mathrm{yr}^{-1}$, which leads to an accretion rate of 
about $4 \times 10^{-6}~M_\odot~\mathrm{yr}^{-1}$ assuming that the accretion rate
is 10 times higher than the jet mass loading rate (e.g. \citealt[]{Ellerbroek13} and references therein). 
The envelope infall rate estimated above is high enough to feed this accretion rate. 

\subsection{Rotating Outflow}
\label{sec:rotationoutflow}

Figure \ref{fig:pv_CS} shows that in the velocity range from 1 to $3.5~\kms$ the
CS (2-1) emission appears to trace a collimated structure along the axis of the redshifted outflow (top panel). 
This jet-like structure starts at the position of the central source and extends to about $50\arcsec$
(22000 AU) along the outflow axis where most of the $\twco$ emission inside the 
outflow cavity ends with the R1 clump and a bright $\Htwo$ knot.
The PV diagram along the axis (Figure \ref{fig:pv_CS}, middle panel) 
shows that the velocity increases with the distance to the central source,
indicating jet bow-shock entrainment as discussed in Section \ref{sec:outflow}.
The coincidence of the CS and $\twco$ emission suggests that
CS is tracing the same material entrained by the jet bow-shock which has 
an apex at the $\twco$ clump R1.

The PV diagrams perpendicular to the outflow axis 
(Figure \ref{fig:pv_CS}, bottom panel) show that there are consistent
velocity gradients perpendicular to the outflow axis from 8$\arcsec$ to $32\arcsec$ 
with respect to the central source and the gradient is highest at about $16\arcsec$ from the central source. 
One possible explanation of this gradient is outflow rotation, considering the direction of the velocity gradient
is consistent with the rotation of the flattened envelope around the central source (see Section \ref{sec:rotationcore}).
We admit that this is only tentative evidence of outflow rotation.
Another possility is that the emission is part of ring-like structures on PV diagrams which can be
produced by the jet bow-shock entrainment (\citealt[]{Lee00}).
However, in such case the highest velocity gradient is expected to be at the apex of the jet bow-shock,
which is not seen here.
The other possibilities include asymmetric shock interaction or jet precession.

If outflow rotation is the case, the projected rotational velocity is 0.4 to $1.1~\kms$ 
with a mean value of 0.7 $\kms$
(the velocity differences between the ends and the mid-points of the red lines in Figure \ref{fig:pv_CS})
at a radius of about $4\arcsec$ ($1800$ AU) from the outflow axis 
(the distance between the ends of the red lines and the outflow axis shown in Figure \ref{fig:pv_CS}).
The specific angular momentum is then $j_\mathrm{CS} \approx 1600~\aukms$ using the above radius and mean
rotational velocity (0.9 $\kms$ after correcting for the inclination of $40^\circ$ of the outflow).
Assuming the transport of angular momentum from a
jet to its entrained outflow is on the same level as the jet's linear momentum
transport, the jet that has entrained the CS outflow 
should have a specific angular momentum
\begin{equation}
j_w=j_\mathrm{CS}\left(\frac{v_{p,w}}{v_{p,\mathrm{CS}}}\right) 
\approx 1600~\left(\frac{v_{p,w}}{v_{p,\mathrm{CS}}}\right)~\aukms,\label{eq:jw}
\end{equation}
where $v_{p,w}$ is the poloidal velocity of the jet,
and $v_{p,\mathrm{CS}}$ is the poloidal velocity of the CS outflow (around $3.5~\kms$
after correcting for the inclination).
We then can estimate the launching radius of this wind 
using the formula provided by \citet[]{Anderson03} 
(Eq. 5 in their paper),
\begin{eqnarray}
\varpi_0 & = &0.7~\mathrm{AU}~\left(\frac{j_w}{100~\aukms}\right)^{2/3}\nonumber\\
& & \left(\frac{v_{p,w}}{100~\kms}\right)^{-4/3}\left(\frac{m_*}{1~M_\odot}\right)^{1/3},
\end{eqnarray}
which combined with Equation \ref{eq:jw} yields
\begin{equation}
\varpi_0 = 62~\mathrm{AU}~\left(\frac{v_{p,w}}{30~\kms}\right)^{-2/3}\left(\frac{m_*}{0.3~M_\odot}\right)^{1/3}.
\end{equation}
We can further deduce the magnetic lever arm following \citet[]{Ferreira06} (Eq. 10 in their paper) to be
\begin{equation}
\lambda=106~\left(\frac{v_{p,w}}{30~\kms}\right)^{4/3}\left(\frac{m_*}{0.3~M_\odot}\right)^{-2/3},
\end{equation}
and the Alfv\'en radius to be
\begin{equation}
\varpi_A=\varpi_0\sqrt{\lambda}=640~\mathrm{AU},
\end{equation}
which is independent of $v_{p,w}$ or $m_*$.
Note that we leave the jet velocity $v_{p,w}$ as a free parameter because 
even though the optical jet reaches about 300~$\kms$,
the CS outflow may be entrained (toroidally) by the slower and less dense part of the jet which is not
seen in optical lines.

Compared with other observations, such as Class 0 molecular jets 
(e.g. \citealt[]{Lee08}; \citealt[]{Lee09}; \citealt[]{Choi11}), Class I $\Htwo$ jets (e.g. \citealt[]{Chrysostomou08}) 
and optical T Tauri jets (e.g. \citealt[]{Coffey07}),
we have detected a similar rotational velocity at a much larger radius, and therefore
higher specific angular momentum. 
Unlike prior observations, we believe here CS is not tracing the material directly launched from the disk
but instead the entrained material.
The detected velocity gradient could be evidence that the CS outflow is entrained not only 
poloidally but also toroidally by a MHD disk wind, and the derived footpoint radius suggests 
that this disk wind is launched from relatively large radii.
We note that the measured specific angular momentum is only an upper limit because a similar
velocity gradient with a smaller resolution beam will give a much smaller
specific angular momentum. In fact, in the literature listed above, the 
reported specific angular momenta tend to be smaller with a higher angular resolution.

\section{Summary and Conclusions}
\label{sec:summary}

We present ALMA Cycle 1 observations of the HH 46/47 molecular outflow,
combining the 12m array, 7m array and single dish total power data.
Compared with previous cycle 0 observations, the new observations have
higher angular resolution ($1.3\arcsec$, nearly 3 times higher than before),
ability to recover extended emission, and
coverage of more molecular species, including $\thco$, $\ceio$, $\cseo$, CS and $\cths$. 
Our main conclusions are as follows.

1) We detect an extended component in the continuum emission, which is elongated
with its major axis perpendicular to the outflow axis. 
Its morphology appears to be affected by the outflow.
We conclude it traces a flattened envelope that is shaped by the bipolar outflow.

2) The new $\thco$ and $\ceio$ data allowed us to trace outflow material with higher column 
density than $\twco$. They are only detected within about $1 - 2~\kms$ from 
the cloud velocity, tracing the outflow to lower velocities than what is possible 
using only the $\twco$ emission. Interestingly, the cavity wall of the red-shifted outflow 
appears at very low velocities (as low as ~0.2 km/s) in emission of these molecules. 

3) We used the $\thco$ and $\ceio$ emission to correct for the CO optical depth and 
accurately estimated the mass, momentum and kinetic energy of the outflow.
Correcting for the $\twco$ optical depth increases the estimated mass of the outflow 
by a factor of 8.5, the momentum by a factor of 4.9, and the kinetic energy by a factor of 2.4. 
Adding the slower material traced only by $\thco$ and $\ceio$,
there is another factor of 3 increase in the mass estimate and 50\% increase in the momentum estimate.
Assuming $\Tex=15$ K, the measured total mass of the outflow is 1.6 $M_\odot$, 
the total momentum is 1.9 $M_\odot~\kms$ (after correcting for inclination) and the total kinetic energy is
$3.9\times 10^{43}$ erg. The estimated outflow mass and momentum are significantly higher
than those previously reported from surveys of Class 0 and I outflows.

4) We derived the spatial distributions of the mass, momentum and kinetic energy of the outflow.
Despite the very different sizes and morphologies of the blue-shifted and red-shifted outflows,
their energy distributions appear to be symmetric, 
and both are concentrated on the outflow cavity walls near the central source. 
Interestingly, even though the jet bow-shocks entrainment is significant in the red lobe, 
more outflow energy is being deposited into the cloud at the base of the outflow cavity rather than 
close to the heads of the bow shocks.

5) By comparing the mass distributions of the outflow and the remaining core,
we find that the molecular outflow is mainly 
composed of locally entrained core material, rather than being made of material that was 
entrained close to the source and then carried out to its current position.
This indicates that the 
core material joins the outflow as the outflow cavity broadens. Based on such a
scenario, we find that the outflow is capable of dispersing the core within
the lifetime of the embedded phase of a typical low-mass protostar.
We also estimated the current instantaneous core-to-star efficiency to be about 1/3,
and the current average core-to-star efficiency to be 1/3 to 1/4, suggesting the outflow
has already been significantly affecting the star formation efficiency.

6) The improved angular resolution and sensitivity allow us to see richer details of the 
outflow structure. Notably, we find that the outflow cavity wall of the red-shifted outflow is composed 
of two or more layers of outflowing gas, which connect with different shocked regions 
along the outflow axis inside the cavity, suggesting that the outflow cavity wall is 
made of multiple shells entrained in a series of jet bow-shock events. 
For the blue-shifted outflow, we showed the CO emission above about $6~\kms$
can be well fitted with the wide-angle wind entrainment model. 
However, we find evidence that both mechanisms are actually in action on both sides of the outflow,
even though the mechanism that is more visible differs on the two sides 
due to the different environment of the outflow.

7) We identify a flattened structure around the central source perpendicular to the outflow
axis in the $\thco$ and $\ceio$ (1-0) maps. The morphologies of this structure
in the $\thco$ and $\ceio$ integrated images indicate that it is shaped by the outflow
cavities. 
Comparison between the observed PV diagrams in both species and a simple analytic model suggest that 
the observed flattened structure can be explained by a collapsing envelope with rotation.
We estimated an envelope infall rate of $3.2 \times 10^{-6}~M_\odot~\yr^{-1}$, which is 
enough to sustain the disk accretion rate suggested by the mass flux of the atomic jet.
Higher angular resolution observations are needed to probe the transition from the infalling envelope
to a rotationally supported disk and even higher resolution is needed to resolve the disks feeding each protostar
in the binary system.

8) At outflow velocities from 1 to $4~\kms$, the CS (2-1) emission traces a collimated structure along the 
outflow axis inside the outflow cavity. Its kinematics and spatial overlap with the $\twco$ emission 
inside the cavity suggest that it is tracing jet-entrained material. We detect
velocity gradients across its axis over its length. If this is due to the
rotation of the outflow, the estimated specific angular momentum is about $1600~\aukms$
and would also imply that the CS outflow is entrained,
not only poloidally but also toroidally, by a disk wind launched from relatively large radii ($60~\mathrm{AU}$).

\acknowledgements

We thank the anonymous referee for helpful discussions. 
This paper makes use of the following ALMA data: ADS/JAO.ALMA \#2012.1.00382.S. 
ALMA is a partnership of ESO (representing its member states), NSF (USA) and NINS (Japan), 
together with NRC (Canada), NSC and ASIAA (Taiwan), and KASI (Republic of Korea), 
in cooperation with the Republic of Chile. The Joint ALMA Observatory is operated by ESO, AUI/NRAO and NAOJ.
 
\appendix

\section{Column Density and Optical Depth Correction}
\label{app:mass}

The radiative transfer equation in the form of radiation temperature is (e.g. \citealt[]{Bourke97})
\begin{equation}
T_R(v)=f\left[J_\nu(\Tex)-J_\nu(\Tbg)\right]\left(1-\exp(-\tau_v)\right),\label{eq:rt}
\end{equation}
where $T_R$ is background-subtracted radiation temperature, $\Tex$ is the excitation temperature 
(assumed constant along the line of sight),
$\Tbg$ is the background temperature,
$f$ is the beam filling factor, and
\begin{equation}
J_\nu(T)\equiv\frac{h\nu/k}{\exp\left(\frac{h\nu}{kT}\right)-1}.
\end{equation}
In local thermodynamic equilibrium (LTE), the optical depth of a transition at velocity $v$ 
relates to the column density of the molecule at that velocity by
\begin{equation}
\frac{dN}{dv}=\left(\frac{8\pi k\nu_{ul}^2}{h c^3 A_{ul} g_u}\right)
\Qrot(\Tex) \exp\left(\frac{E_u}{k\Tex}\right) J_{\nu_{ul}}(\Tex) \tau_v,\label{eq:tau}
\end{equation}
where $\nu_{ul}$ is the frequency of the transition, $A_{ul}$ is the
Einstein A coefficient, $E_u$ and $g_u$ are the energy and degeneracy of the upper level, 
and $\Qrot$ is the partition function ($\Qrot (T)=\sum^\infty_{J=0}g_J e^{-E_J/kT}
=\sum^\infty_{J=0}(2J+1) e^{-B_0J(J+1)/kT}$).
We can estimate the column density from the measured intensity 
by combining Eqs. \ref{eq:rt} and \ref{eq:tau},
\begin{equation}
\frac{dN}{dv}=\left(\frac{8\pi k\nu_{ul}^2}{h c^3 A_{ul} g_u}\right)
 \Qrot(\Tex) \exp\left(\frac{E_u}{k\Tex}\right)
\frac{J_{\nu_{ul}}(\Tex) }{[J_\nu(\Tex)-J_\nu(\Tbg)]}
\left(\frac{\tau_v}{1-\exp(-\tau_v)}\right) \frac{T_R(v)}{f}.
\end{equation}
Assuming $\Tbg=2.7$~K, $J(\Tbg)$ is insignificant compared to $J(\Tex)$
for typical $\Tex > 10$~K at a frequency around 100 GHz. Also with a line width
of $\lesssim 100~\kms$, $J_\nu(\Tex)\approx J_{\nu_{ul}}(\Tex)$.
Therefore we have
\begin{equation}
\frac{dN}{dv}=\left(\frac{8\pi k\nu_{ul}^2}{h c^3 A_{ul} g_u}\right)
 \Qrot(\Tex) \exp\left(\frac{E_u}{k\Tex}\right)
\left(\frac{\tau_v}{1-\exp(-\tau_v)}\right) \frac{T_R(v)}{f}.\label{eq:dndv}
\end{equation}
In the optically thin limit, $\tau\ll 1$, $\tau_v/(1-\exp(-\tau_v)) \approx 1$, and
\begin{equation}
\frac{dN}{dv}= \left(\frac{8\pi k\nu_{ul}^2}{h c^3 A_{ul} g_u}\right)
 \Qrot(\Tex) \exp\left(\frac{E_u}{k\Tex}\right)\frac{T_R(v)}{f}.
\end{equation}
Therefore
\begin{equation}
\frac{dN}{dv}=\frac{dN}{dv}\bigg|_{\mathrm{thin}}F_\tau(v),
\end{equation}
with the optical depth correction factor 
\begin{equation}
F_\tau(v)\equiv \frac{\tau_v}{1-\exp(-\tau_v)}.
\end{equation}
For the $\twco$ (1-0) line, we adopt $\nu_{ul}=115.271$ GHz, $A_{ul}=7.203 \times 10^{-8}~\mathrm{s}^{-1}$,
$g_u=2J_u+1=3$, $E_u=5.53$ K, $B_0/k=2.765$ K. 
We adopt $\nu_{ul}=110.201$ GHz, $A_{ul}=6.294 \times 10^{-8}~\mathrm{s}^{-1}$,
$g_u=2J_u+1=3$, $E_u=5.29$ K, $B_0/k=2.645$ K for the $\thco$ (1-0) line, 
and $\nu_{ul}=109.782$ GHz, $A_{ul}=6.266 \times 10^{-8}~\mathrm{s}^{-1}$,
$g_u=2J_u+1=3$, $E_u=5.27$ K, $B_0/k=2.635$ K for the $\ceio$ (1-0) line.

We estimate the optical depth correction factor $F_\tau$ following the method outlined
by \citet[]{Dunham14}. Assuming $\twco$ and $\thco$ trace the same material and have the same 
$\Tex$ and $f$, from Eq. \ref{eq:rt} we have
\begin{equation}
\frac{T_{R,12}(v)}{T_{R,13}(v)}=\frac{1-\exp(-\tau_{v,12})}{1-\exp(-\tau_{v,13})},\label{eq:tratio}
\end{equation}
where the subscripts 12 and 13 represent $\twco$ and $\thco$ respectively.
If $\thco$ is optically thin ($\tau_{13} \ll 1$),
\begin{equation}
\frac{T_{R,12}(v)}{T_{R,13}(v)}
=\frac{1-\exp(-\tau_{v,12})}{\tau_{v,13}}\approx X_{12,13}\frac{1-\exp(-\tau_{v,12})}{\tau_{v,12}},
\end{equation}
where $X_{12,13}$ is the abundance ratio between $\twco$ and $\thco$.
The last step is valid because the same transitions of the isotopologues at the same
excitation temperature have very similar $\nu_{ul}$, $A_{ul}$, $E_u$, $g_u$, $\Qrot(\Tex)$
and then from Eq. \ref{eq:tau} we have 
\begin{equation}
\frac{\tau_{12}}{\tau_{13}}\approx\frac{dN_{12}/dv}{dN_{13}/dv}=X_{12,13}.
\end{equation}
Therefore the correction factor for the $\twco$ optical depth can be estimated as
\begin{equation}
F_{\tau,12}(v)=X_{12,13}\frac{T_{R,13}(v)}{T_{R,12}(v)},
\end{equation}
assuming $\thco$ is optically thin.

$\thco$ may not be optically thin at low velocities and we can use
the less abundant and more optically thin isotopologue $\ceio$ to correct the optical depth of 
$\thco$. In such a case, we have
\begin{equation}
F_{\tau,13}(v)=X_{13,18}\frac{T_{R,18}(v)}{T_{R,13}(v)},
\end{equation}
where the subscript 18 represents $\ceio$, $X_{13,18}$ is the abundance ratio between $\thco$ and $\ceio$,
and
\begin{equation}
F_{\tau,12}(v)=X_{12,13}\frac{T_{R,13}(v)F_{\tau,13}}{T_{R,12}(v)}=X_{12,13}\frac{T'_{R,13}(v)}{T_{R,12}(v)},
\end{equation}
where $T'_R\equiv T_{R}(v)F_{\tau}$ is optical-depth-corrected intensity.
If the two isotopologues used for intensity ratios are both optically thin, we have 
$T_{R,12}/T'_{R,13}=X_{12,13}$ and $T_{R,13}/T_{R,18}=X_{13,18}$ as their upper limits, i.e., a
lower limit of 1 for the optical depth correction factor $F_\tau$.
We note again that this method assumes that these isotopologues trace the same material 
at the same excitation
temperature under LTE conditions and have constant abundance ratios.

\end{document}